\documentclass[sigconf]{acmart}
\AtBeginDocument{%
  }

\setcopyright{none}

\usepackage{listings}
\usepackage{color-edits}
\addauthor{todo}{red}

\begin{document}

\title{EvolveCaptions: Supporting DHH Users Through Collaborative Error-Driven ASR Personalization}


\author{Liang-Yuan Wu}
\affiliation{%
  \institution{University of Michigan}
  \city{Ann Arbor, MI}
  \country{USA}}
\email{lyuanwu@umich.edu}

\author{Dhruv Jain}
\affiliation{%
  \institution{University of Michigan}
  \city{Ann Arbor, MI}
  \country{USA}}
\email{profdj@umich.edu}

\begin{abstract}
Automatic Speech Recognition (ASR) systems often struggle to accurately transcribe speech from Deaf and Hard of Hearing (DHH) speakers. Existing personalization approaches typically require substantial pre-recorded datasets and burden DHH users. We present EvolveCaptions, a collaborative, error-driven personalization system that connects caption correction with speaker-specific ASR adaptation. When transcribing a DHH speaker's speech, hearing partners repair transcription errors, and the system generates short, targeted clauses for the DHH speaker to record. These samples are used to incrementally fine-tune the model. In a study with 12 DHH and six hearing participants, EvolveCaptions reduced Word Error Rate (WER) across five scripted  reading rounds, requiring an average of five minutes of additional speech recording. Participants found the interfaces usable but considered recording and correction effortful; acceptability depended on context and anticipated long-term benefit. These findings show the potential of collaborative, cumulative adaptation while exposing the labor and conditions needed to sustain it.
\end{abstract}



\keywords{Accessibility, Deaf and Hard of Hearing, Automatic Speech Recognition.}
\begin{teaserfigure}
  \includegraphics[
    width=\textwidth,
    alt={Three-stage cycle connecting correction, targeted recording, and speaker-specific automatic speech recognition adaptation. In Stage 1, a Deaf or Hard of Hearing (DHH) speaker's live caption incorrectly reads ``get hub,'' and a hearing partner changes it to ``GitHub.'' The corrected term is sent to a large language model. In Stage 2, the model generates the clause ``I will push the code to GitHub today,'' which the DHH speaker records. In Stage 3, the recording and transcript fine-tune the speaker's automatic speech recognition model. The updated model loops back to caption the speaker's future speech.}
  ]{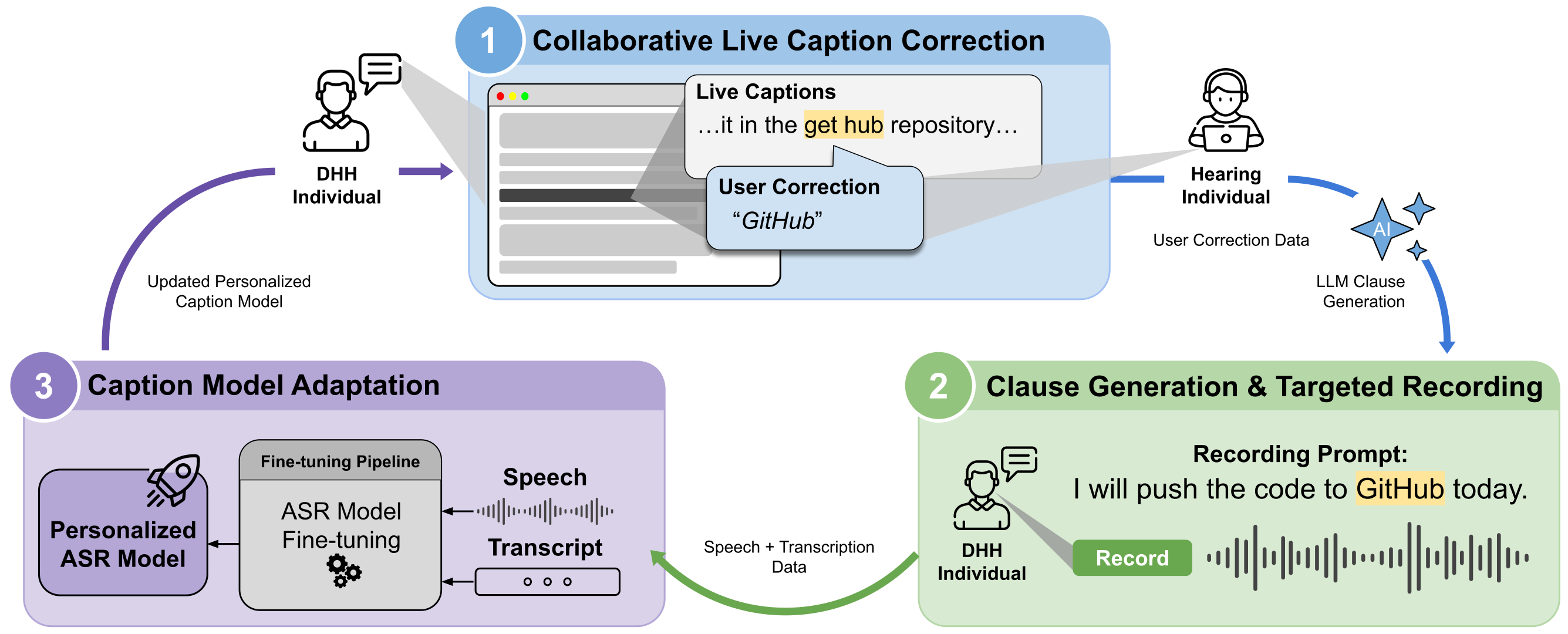}
  \caption{\textit{EvolveCaptions} is an interactive ASR adaptation system that improves accessibility in conversations involving DHH and hearing participants. It combines live caption correction with iterative, speaker-specific fine-tuning: (1) Hearing users correct live captions of the DHH speaker’s voice; (2) the DHH speaker records targeted clauses generated from the corrected terms; (3) the Whisper ASR model is fine-tuned with these recordings and adapts to the speaker over time.}
  \Description{Three-stage cycle connecting correction, targeted recording, and speaker-specific automatic speech recognition adaptation. In Stage 1, a Deaf or Hard of Hearing (DHH) speaker's live caption incorrectly reads ``get hub,'' and a hearing partner changes it to ``GitHub.'' The corrected term is sent to a large language model. In Stage 2, the model generates the clause ``I will push the code to GitHub today,'' which the DHH speaker records. In Stage 3, the recording and transcript fine-tune the speaker's automatic speech recognition model. The updated model loops back to caption the speaker's future speech.}
  \label{fig:teaser}
\end{teaserfigure}


\maketitle

\section{Introduction}
Automatic Speech Recognition (ASR) has become a widely used technology for enhancing communication. Yet, the diversity of speaking styles presents a persistent challenge to developing universally accurate ASR systems capable of understanding all users \cite{cerva2012real, berke2017deaf, rowe2022characterizing}. While state-of-the-art models demonstrate impressive performance on benchmark datasets \cite{ghimire2024comprehensive}, they continue to underperform on speech produced by Deaf and Hard of Hearing (DHH) individuals \cite{fok2018towards, zhao2025quantification}. Speech among DHH speakers is heterogeneous, with variations in pronunciation, rhythm, and articulation that may affect intelligibility for both ASR systems and human listeners, including familiar communication partners \cite{mattys2013speech, bigham2017deaf}. Because these speech patterns remain underrepresented in ASR training data, recognition errors can disproportionately burden DHH speakers and contribute to inequities in communication access \cite{elliot2017user, glasser2017deaf}.

Prior efforts to improve ASR for DHH users, such as Project Euphonia \cite{martin2025project} and the work of Tobin et al. \cite{tobin2022personalized}, have focused on personalization through large, pre-recorded datasets collected in controlled conditions. However, these approaches face three key limitations. First, they impose a substantial motivational burden: spending hours recording scripted speech is often impractical for users. Second, they lack contextual relevance: samples collected offline may not generalize to spontaneous, real-world interaction. Third, they frame adaptation as an isolated, pre-emptive task, placing the full responsibility on the DHH individual without opportunities for real-time, shared refinement.

In this paper, we introduce \textit{EvolveCaptions}, a collaborative, error-driven ASR adaptation system that connects live caption correction with speaker-specific adaptation during interaction between DHH speakers and hearing communication partners. The system transcribes speech in real time while hearing partners review and correct errors. These corrections guide the generation of short, targeted clauses, which the DHH speaker records to incrementally fine-tune the ASR model. This loop repeats iteratively, enabling the system to adapt over time without requiring extensive pre-collected training data.

By deriving prompts from observed misrecognitions, EvolveCaptions grounds data collection in errors encountered during use. The workflow collects adaptation data incrementally, while hearing partners contribute corrections and DHH speakers control whether the resulting prompts are recorded. This design aligns with frameworks of collaborative accessibility and interdependence, which emphasize shared responsibility for access rather than individual accommodation \cite{bennett2018interdependence, branham2015collaborative}.

We evaluated EvolveCaptions in a structured remote study with 12 DHH participants and six hearing partners. Across five sequential  captioning rounds, DHH participants read randomized scripts while hearing partners corrected captioning errors in real time. These corrections generated targeted prompts for the DHH participants to record, enabling incremental model adaptation between rounds.

Comparing the first and final rounds, EvolveCaptions reduced word error rate (WER) by a median of 27.2\% (mean = 30.4\%), a statistically significant improvement ($p < .05$, Wilcoxon signed-rank test). To contextualize these gains, we compared the final personalized models with two fixed ASR baselines: an unmodified Whisper-base model and a Whisper-base model generically fine-tuned on dysarthric speech. We also included the human-corrected Round 1 transcript as a reference point illustrating the immediate effect of manual correction. By Round 5, EvolveCaptions achieved lower WER than both fixed-model baselines for all participants, providing evidence for the value of iterative, speaker-specific adaptation. Qualitative findings further show that participants experienced the system as shifting accessibility from requiring DHH users to adapt to technology toward collaborative, cumulative system learning, while also highlighting the conditions under which users are willing to engage in such adaptation.

In summary, this work contributes: (1) \textit{EvolveCaptions}, an interactive captioning system that supports collaborative, error-driven ASR personalization; (2) an empirical evaluation demonstrating both quantitative improvements and qualitative insights into how accessibility labor is distributed and negotiated in collaborative captioning; and (3) design implications for future collaborative captioning systems, highlighting how accessibility can be supported as a shared, evolving process grounded in user agency, trust, and context. We release the project data and artifacts through:~\url{https://anonymous.4open.science/r/EvolveCaptions-Anonymized}.

\section{Related Work}
We examine DHH captioning needs and ASR’s role in communication between DHH and hearing participants, then situate our work within three areas: (1) ASR-based captioning for DHH users, (2) personalized ASR for atypical speech, and (3) collaborative captioning systems that share accessibility labor.

\subsection{Captioning Needs of DHH People}

The DHH community encompasses a diverse population with varying degrees of hearing loss, communication preferences, and cultural identities \cite{cavender2008hearing}. Some individuals identify as culturally Deaf, using American Sign Language (ASL) as their primary language and participating in a shared cultural-linguistic community \cite{moore1993hearing, ladd2003understanding, cavender2008hearing}. Others may identify as hard of hearing or audiologically deaf, often using spoken language and relying on hearing technologies such as hearing aids or cochlear implants \cite{cavender2008hearing, ladd2003understanding}. Communication access needs across this spectrum are highly individualized, and no single approach suffices for all users.

Captions serve as a crucial accessibility tool for many DHH individuals, enabling access to spoken content across educational, professional, and social domains \cite{millett2021accuracy, iezzoni2004communicating, jain2018exploring, loizides2020breaking, kushalnagar2013captions, kushalnagar2020teleconference}. Prior work has examined how DHH users adopt captioning technologies in classrooms \cite{millett2021accuracy, kawas2016improving, kushalnagar2014accessibility}, medical settings \cite{iezzoni2004communicating, hughes2026assessing}, workplaces \cite{jain2018exploring, elliot2016deaf}, and daily communication contexts \cite{mcdonnell2022understanding, prietch2014speech}. These studies emphasize that captions not only improve comprehension, but also support participation, independence, and equitable engagement.

Several modalities exist for caption provision. Human-generated captions, such as those provided by Communication Access Realtime Translation (CART) professionals, offer high accuracy but are expensive and not easily scalable \cite{cartoverview}. Crowd-sourced and peer-generated captions have been explored as lower-cost alternatives with mixed results \cite{harrington2013crowd, lasecki2012real}. More recently, Automatic Speech Recognition (ASR) has become a popular method for generating real-time captions due to its accessibility, scalability, and low latency \cite{seita2022remotely, kafle2017evaluating, wald2008universal}. However, ASR captions often fall short in accuracy when applied to DHH speech, leading to breakdowns in communication and increased cognitive burden on users \cite{zhao2025quantification, kafle2017evaluating}.

\subsection{ASR for DHH--Hearing Communication}

ASR is increasingly used to facilitate DHH--hearing communication in classrooms, meetings, and daily interactions \cite{ seita2022remotely, elliot2017user, yamamoto2021see, chen2024towards}. Tools like Google Live Transcribe \cite{google_transcribe}, Otter.ai \cite{otter_ai}, and Zoom's live captions \cite{zoom_captions}, as well as open-source models like Whisper \cite{radford2023robust}, offer low-cost, real-time transcription with broad accessibility.

Despite these advances, ASR systems consistently underperform on speech from DHH users. Most commercial ASR models are trained on fluent, non-disabled speech \cite{shor2019personalizing, tobin2022personalized, rudzicz2010towards}, resulting in poor performance when processing speech that diverges in rhythm, articulation, and prosody. This underrepresentation can substantially degrade ASR accuracy \cite{zhao2025quantification, pradhan2018accessibility, mengistu2011comparing}. Additionally, misalignments between how DHH speakers monitor or produce speech and how ASR systems interpret it may further reduce reliability \cite{mattys2013speech, moore1993hearing, de2010longitudinal, rodolitz2019accessibility}.

These transcription failures can reduce intelligibility and erode user trust, especially in communication involving both DHH and hearing participants where mutual understanding is critical \cite{kafle2017evaluating, gottermeier2016user, berke2017deaf}. Common ASR issues such as high Word Error Rates (WER), hallucinated phrases, and missing function words can significantly distort meaning, further increasing the cognitive burden on DHH users. While Whisper and similar models perform well on standard benchmarks \cite{radford2023robust}, their robustness does not consistently extend to speakers with atypical speech patterns \cite{radford2023robust, tobin2022personalized}. These gaps underscore the need for adaptive ASR systems that can provide equitable communication access across diverse speakers.

\subsection{Personalized ASR for Speaker Variation}

Personalized ASR \cite{baskar2022speaker} seeks to improve accuracy by adapting models to individual speakers, an approach which has been shown to be effective for people with disordered or atypical speech \cite{shor2019personalizing, tobin2022personalized, rudzicz2010towards, macdonald2021disordered, jiang2024perceiver}. In DHH contexts, where speech can vary widely in rhythm, clarity, and articulation, personalization has the potential to significantly improve transcription quality \cite{zhao2025quantification}.

Projects such as Google's Project Euphonia \cite{martin2025project} and Project Relate \cite{google_relate} have demonstrated that models fine-tuned on even a few minutes of speaker-specific data can substantially reduce WER. Tobin et al. \cite{tobin2022personalized} found that most participants achieved usable accuracy levels with only 3--4 minutes of training data, while other work has explored low-resource fine-tuning of Whisper for dysarthric speech \cite{mulfari2024voice}. Adjacent HCI work has explored speech restoration rather than transcription personalization: CLARIS converts whispered and dysarthric input into more intelligible speech and supports adaptation to new speakers using 15--30 minutes of speaker-specific data \cite{shah2026claris}. Contextual biasing and speech-based in-context learning can condition recognition on expected phrases or example utterances \cite{huang2023contextualized, wang2024can}, but they do not produce persistent speaker-specific parameter updates.

Recent HCI and deployment work has also brought personalization into user-facing workflows. Adapt4Me presents an uncertainty-aware authoring environment in which users resolve uncertain ASR outputs and incrementally personalize the recognizer \cite{pokel2026demonstration}. In a single-speaker deployment, Huber et al. incorporated transcript corrections into subsequent fine-tuning for a dysarthric speaker \cite{huber2026adapting}. Together, these systems establish that interactive and correction-derived personalization are possible. However, they do not examine how this work is divided between a DHH speaker and a communication partner.

\subsection{Collaborative Captioning}

While much of the accessibility literature focuses on individual accommodations, recent HCI research has emphasized \textit{collaborative accessibility} \cite{branham2015collaborative}---designing systems where access is a shared responsibility \cite{seita2022remotely, mcdonnell2022understanding, mcdonnell2023easier}. This perspective, grounded in the principle of \textit{collective access}, highlights the role of interdependence in accessibility relationships between DHH speakers and their communication partners \cite{bennett2018interdependence}.

Several systems have explored how hearing users can assist in real-time caption correction. For example, McDonnell et al. \cite{mcdonnell2022understanding} examined how hearing participants support caption repair in Zoom meetings. Others have enabled crowd-sourced or partner-based corrections of ASR transcripts to improve live communication \cite{harrington2013crowd,fok2018towards,kuhn2024record}. More recently, Kuhn et al. evaluated real-time ASR correction by non-professionals, finding that accuracy improved while identifying correction as a demanding task \cite{kuhn2025communication}.

To reduce the continuous manual correction effort, a growing body of work explores automated transcript refinement. CARTGPT leverages LLMs to refine professional CART captions \cite{wu2025cartgpt}, while frameworks like LIR-ASR generate phonetically similar variants to correct standard ASR outputs \cite{liu_listening_2025}. Hernandez et al. incorporate word-level confidence into LLM-based correction for disordered speech, directing changes toward uncertain regions and reducing overcorrection \cite{hernandez2026confidence}. These approaches improve the current transcript through post-processing, but do not update a speaker-specific ASR model.

Our work connects collaborative caption correction with subsequent speaker-specific adaptation. A hearing corrector's edits or uncertainty markings identify candidate errors, while the DHH speaker provides recording samples derived from those ASR errors. The corrections therefore focus the collection of speaker-specific training data rather than being treated automatically as reliable ground-truth labels. The resulting speech--text pairs update the model for subsequent captioning. Building on design paradigms that center DHH individuals within shared auditory interfaces \cite{wu2025soundnarratives}, we examine how this workflow distributes effort and control between DHH speakers and hearing partners.

\section{EvolveCaptions: A Collaborative Error-Driven ASR Personalization System}
\textit{EvolveCaptions} is a collaborative, error-driven ASR personalization system designed to improve recognition of speech from DHH speakers in communication with hearing partners.  The system displays live captions of a DHH speaker’s speech and allows a hearing corrector to edit recognition errors. It converts corrected terms into short recording prompts, which the DHH speaker can review, record, skip, or rerecord. The resulting speech–text pairs are used to update a speaker-specific ASR model between interaction rounds, and the updated model is used for subsequent captioning. Thus, correction occurs live, while personalization proceeds iteratively from errors encountered during use. The system is designed for use in recurring communication settings, where speaker-specific captioning improvements can accumulate across interactions. Below, we describe the design motivations, interaction workflow, and technical implementation of the system.

\subsection{Design Goals}
\label{sec:design}

EvolveCaptions is grounded in three design goals, informed by prior work in ASR personalization, accessibility, and collaborative interaction:

\begin{enumerate}
    \item \textbf{Minimize additional personalization effort.} Speaker-specific ASR personalization should limit the need for a separate, extensive enrollment process. Rather than asking speakers to anticipate and record a large generic corpus~\cite{tobin2022personalized, shor2019personalizing}, personalization should prioritize speech examples relevant to recognition difficulties encountered by the speaker.
    
    \item \textbf{Ground personalization in errors encountered during use.} Personalization should draw on recognition failures produced by the target speaker’s own speech rather than rely exclusively on generic, predetermined enrollment content. These failures indicate where a general-purpose ASR model has difficulty recognizing that speaker and provide speaker-relevant targets for collecting additional speech examples. Corrections to these errors should inform subsequent speaker-specific recognition rather than benefit only the current transcript \cite{zhao2025quantification}.

    \item \textbf{Support collective access through collaborative correction.} Communication access should not depend solely on DHH speakers adapting themselves to recognition failures. Drawing on collective-access approaches~\cite{seita2022remotely, mcdonnell2022understanding}, communication partners should be able to assist with identifying caption errors, while DHH speakers retain authority over their intended meaning and contributed speech data.
\end{enumerate}

\subsection{Three-Stage Interaction Loop}

To meet the above design goals, EvolveCaptions follows a three-stage interactive loop. First, a hearing corrector identifies errors in the DHH speaker’s live ASR output. Second, the system converts corrected terms into short recording prompts that the DHH speaker can review, record, skip, or rerecord. Third, the accepted recordings are paired with their prompt text and used to update the speaker-specific ASR model between interaction rounds. Correction therefore occurs live, while model personalization occurs after recording and before subsequent captioning
 (Figure \ref{fig:teaser}).

\subsubsection{Collaborative Live Caption Correction}

The system transcribes the DHH speaker’s voice using a Whisper-based ASR engine and displays real-time captions to all participants. Hearing users can collaboratively correct transcription errors by selecting and editing individual words or short phrases (the first panel in Figure \ref{fig:teaser}). Because hearing users share the immediate utterance context, their live corrections provide a human interpretation of recognition errors, reducing reliance on automated correction and associated hallucination risks. Drawing on prior crowd-correction interfaces \cite{harrington2013crowd, fok2018towards}, our interface allows users to both mark corrections (highlighted in yellow) and flag uncertain segments (in red). This dual-marking mechanism distinguishes between a proposed correction and an uncertainty marking, allowing correctors to indicate that a segment may be incorrect without supplying a replacement. Proposed changes are synchronized to the DHH speaker’s interface, where the speaker can approve a correction, revise it further, or revert to the original ASR output. The DHH speaker retains final authority over the wording used for personalization.

\subsubsection{Clause Generation and Targeted Recording}

Once a caption has been corrected, the system uses the revised word(s) to generate a clause for the DHH speaker to record. Rather than prompting users to repeat isolated words, EvolveCaptions embeds the corrected term into a short clause intended to provide a natural spoken context for recording (the second panel in Figure \ref{fig:teaser}).

To achieve this, we prompt OpenAI GPT-4 with the following instruction:

\begin{lstlisting} 
You are generating short, spoken English clauses to help improve an automatic speech recognition (ASR) system. Based on a word that was misrecognized by ASR, your goal is to create a new clause (5-15 words) that:

--- Sounds natural in a daily conversation 
--- Contains the corrected word in a prominent, clear context 
--- Has a similar phonetic structure to the original sentence 

Original words: "{original}" 
Corrected words: "{corrected}"

Generate one new clause that can be used to help the ASR model learn this correction. Just reply with the clause (no quotes, no explanation). 
\end{lstlisting}

For example, if the ASR originally transcribed \textit{“fok”} and it was corrected to \textit{“fork”}, the system might generate: \textit{“She picked up the fork from the table.”} These targeted, phonetically similar clauses provide a sentence-level recording context and are designed to support effective fine-tuning. These samples focus on previously misrecognized terms and provide targeted speech-text pairs for model adaptation without requiring large amounts of generic speech data.

DHH users are then prompted to record these generated clauses via a recording interface. Each clause includes a waveform display, visual feedback, and playback controls. This supports users who rely on visual feedback and allows them to rerecord, skip, or delete prompts at will. This interaction design gives users control over when and how their voice is used to train the system. By grounding recordings in natural clauses and offering flexible participation, the system collects speaker-produced recordings for model personalization.

\subsubsection{Caption Model Adaptation}

After each recording phase, the accepted clause recordings are formatted into audio–text pairs using the Hugging Face dataset structure. Each sample is padded and collated for batch training, and the ASR model is fine-tuned using Seq2SeqTrainer with lightweight hyperparameters: learning rate of 1e-5, batch size of 8, and a maximum of 100 steps (Figure \ref{fig:teaser}, stage 3). To balance performance and efficiency, we perform full-model fine-tuning on Whisper-base (74M parameters, <1 GB memory). We selected Whisper-base because its size allowed full-model fine-tuning within the between-round interval available in our deployment. Updating all parameters produces a speaker-specific checkpoint for subsequent captioning.

EvolveCaptions maintains a separate model profile for each DHH speaker. The first adaptation round begins from the same checkpoint for every speaker. Each subsequent round begins from that speaker’s checkpoint produced by the previous round. An update uses only the new recordings collected during the immediately preceding recording phase.  After an update completes, the backend loads the resulting speaker-specific checkpoint for captioning in the next reading round. On the NVIDIA T4 GPU used in our study deployment, each model update completed in under one minute.

During inference, the system receives 16kHz, 16-bit PCM audio via WebSocket, performs chunked transcription using the active speaker-specific checkpoint, and streams captions with low latency—comparable to commercial services like Google Live Transcribe \cite{loizides2020breaking}.

\subsection{User Interface}

\begin{figure*}[h]
  \centering
  \includegraphics[
    width=\linewidth,
    alt={Four screenshots show the EvolveCaptions workflow from captioning to model training. Panel 1 shows the inactive Caption tab with instructions and a Start button. Panel 2 shows live captions while captioning is active. Panel 3 shows the hearing corrector's edits: proposed corrections are highlighted in yellow, and uncertain words in red. Panel 4 shows the Record tab with a generated practice phrase, a waveform and duration for one recording, Done and Cancel controls, and a Train button.}
  ]{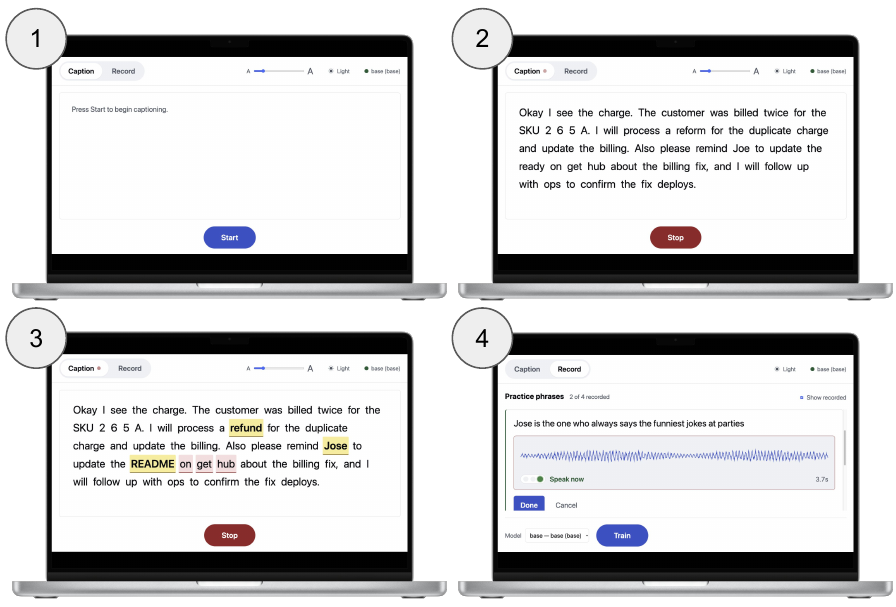}
  \caption{EvolveCaptions user interface. (1) Clicking “Start” in the Caption tab begins generating real-time captions, while the “Record” tab displays a list of targeted clauses for the DHH speaker to record; (2) real-time captions reflect the DHH speaker’s speech; (3) hearing correctors can refine captions by correcting errors (yellow highlights) or flagging uncertain words (red highlights); (4) during recordings, the interface shows targeted samples along with live waveforms visualizing the captured audio signal.}
  \Description{Four screenshots show the EvolveCaptions workflow from captioning to model training. Panel 1 shows the inactive Caption tab with instructions and a Start button. Panel 2 shows live captions while captioning is active. Panel 3 shows the hearing corrector's edits: proposed corrections are highlighted in yellow, and uncertain words in red. Panel 4 shows the Record tab with a generated practice phrase, a waveform and duration for one recording, Done and Cancel controls, and a Train button.}
  \label{fig:ui}
\end{figure*}

EvolveCaptions provides a browser-based interface with two primary modes: live captioning and targeted recording (Figure \ref{fig:ui}). In the captioning mode, selecting “Start ASR” begins streaming the DHH speaker’s audio to the transcription backend and displays the resulting captions to both participants. The hearing corrector can select a recognized word or phrase and either replace it with a proposed correction or mark it as uncertain. Corrected segments are displayed in yellow, while uncertain segments are displayed in red. Changes are synchronized across participants so that both the DHH speaker and hearing corrector can view the revised transcript.

In the recording mode, selecting “Start Recording” presents the DHH speaker with short clauses generated from previously corrected terms. For each clause, the speaker can record an utterance, view a waveform of the captured audio, play back the recording, rerecord it, or skip the prompt. The waveform provides visual feedback about the presence and amplitude of the captured audio; it does not assess pronunciation or recording quality. These controls allow the DHH speaker to determine which proposed clauses become speech contributions for speaker-specific personalization.


\subsection{System Implementation}

EvolveCaptions is implemented as a cross-platform web application. The frontend, built with ReactJS \cite{reactjs} and Vite \cite{vite}, handles real-time caption display, collaborative correction, and DHH audio recording.  Audio input is captured using AudioWorklet \cite{audioworklet} and streamed to the backend over WebSocket. The frontend also synchronizes caption updates and correction events across the DHH speaker and hearing-corrector interfaces.

The FastAPI backend manages transcription, correction tracking, clause generation, and between-round model fine-tuning. We extend the open-source WhisperLive\footnote{https://github.com/collabora/WhisperLive} project to support low-latency, chunked transcription and speaker-specific model profiles. During captioning, the backend transcribes incoming audio using the checkpoint associated with the active DHH speaker. At the end of each recording phase, it prepares the training data, launches fine-tuning, and updates the speaker-specific checkpoint.

For our user study, the system was hosted on a Google Cloud VM with an NVIDIA T4 GPU. Client–server communication used HTTPS and secure WebSocket connections, and the backend used a configured CORS policy to restrict permitted frontend origins. Each DHH participant was assigned a separate model profile, which stored captioning model checkpoints and personal recording data. We also confirmed that the core application could run locally on a MacBook Pro (Intel i7), where the system remained fully functional with latency comparable to common real-time captioning tools (e.g., Google Live Transcribe or Zoom captions), demonstrating portability across environments.

To support reproducibility and future research, the full implementation is released at:~\url{https://anonymous.4open.science/r/EvolveCaptions-Anonymized}.

\section{Evaluation}
We conducted a structured remote study to examine the technical feasibility and participant experience of EvolveCaptions. The study included 12 DHH speakers and six hearing correctors, with each DHH participant paired with one hearing corrector for an approximately 90-minute session. During each session, participants completed five scripted reading rounds with four intervening correction, recording, and model-adaptation phases. The study was designed to address three research questions that examine our design goals (Section \ref{sec:design}): minimizing personalization effort (RQ1 and RQ2), grounding personalization in encountered errors (RQ1), and supporting collective access (RQ2 and RQ3):


\begin{enumerate}
    \item How does speaker-specific ASR performance change across five rounds of collaborative error-driven personalization? 
    \item How do DHH speakers and hearing correctors experience their respective roles in the correction, recording, and adaptation workflow?
    \item What considerations shape participants’ anticipated willingness to use collaborative ASR personalization in future communication settings?
\end{enumerate}

\subsection{Participants}

We recruited 12 DHH participants (5 women, 7 men; see Table \ref{tab:par1}) via email lists, social media, and snowball sampling. Eligibility required participants to identify as Deaf or hard of hearing, use spoken English for communication, and be able to participate in the remote speech-based study. The average age was 37.6 years ($SD=18.8$, $median=30$, $range=25-85$). Ten identified as hard of hearing (HoH), one as deaf, and one as Deaf. Eleven participants used hearing devices---nine used hearing aids and three used cochlear implants (some used both). Nine reported regular use of captioning tools (e.g., Apple CC \cite{apple_cc}, Google Transcribe \cite{google_transcribe}, Zoom Captioning \cite{zoom_captions}, CART \cite{cartoverview}) to support verbal communication.

\begin{table*}[h]
\caption{Demographic and communication characteristics of DHH participants.}
\Description{Demographic table for 12 Deaf and Hard of Hearing participants. Columns give participant ID, age, gender, Deaf or hard-of-hearing identity, self-reported hearing loss, age of onset, and paired hearing-corrector ID. Participants are 25 to 85 years old and include five women and seven men. Ten identify as hard of hearing, one as Deaf, and one as deaf. Hearing loss ranges from mild to profound, and onset ranges from birth to 25 years of age.}
\label{tab:par1} 
\begin{tabular}{lllllll}
\toprule
ID  & Age & Gender & Identity & Self-Reported Hearing Loss      & Age of Onset  & Paired Participant\\ 
\midrule
P1  & 38  & Female & HoH      & Severe            & 18 months   &H1\\
P2  & 25  & Male   & Deaf     & Profound          & 5 years    &H2\\
P3  & 68  & Male   & HoH      & Severe            & Birth      &H2\\
P4  & 30  & Male   & HoH      & Moderately Severe & 25 years   &H3\\
P5  & 85  & Male   & HoH      & Profound          & Birth      &H2\\
P6  & 32  & Female & HoH      & Severe            & Birth      &H3\\
P7  & 29  & Female & HoH      & Mild              & 21 years   &H4\\
P8  & 30  & Male   & HoH      & Moderately Severe & Birth      &H4\\
P9  & 28  & Male   & HoH      & Moderate          & Birth      &H5\\
P10 & 25  & Male   & HoH      & Moderate          & Birth      &H5\\
P11 & 31  & Female & deaf     & Profound          & 1 year     &H6\\
P12 & 30  & Female & HoH      & Moderate          & Birth      &H5\\ 
\bottomrule
\end{tabular}
\end{table*}

To serve as collaborative caption correctors, we also recruited six hearing participants (3 women, 3 men, Table \ref{tab:par2}) via lab mailing lists and social media. Their average age was 25.5 years ($SD=2.9$, $median=25.5$, $range=21-29$). All were proficient in spoken English, including two native speakers, one bilingual speaker, and three fluent second-language speakers. Four participants reported prior experience in communicating with DHH people, while two reported no such experience.

\begin{table}[h]
\caption{Demographic characteristics of hearing participants.}
\Description{Demographic table for six hearing caption correctors. Columns give participant ID, age, gender, and English proficiency. Participants are 21 to 29 years old and include three women and three men. Two are native English speakers, one is bilingual, and three are proficient English speakers.}
\label{tab:par2} 
\begin{tabular}{llllll}
\toprule
ID & Age & Gender & English Proficiency \\
\midrule
H1 & 21  & Female & Native Speaker     \\
H2 & 25  & Female & Bilingual          \\
H3 & 24  & Male   & Native Speaker      \\
H4 & 29  & Male   & Proficient          \\
H5 & 28  & Female & Proficient          \\
H6 & 26  & Male   & Proficient        \\
\bottomrule
\end{tabular}
\end{table}

Each session paired one DHH speaker with one hearing corrector. To facilitate scheduling, H1 and H6 each participated in one session, H3 and H4 in two sessions, and H2 and H5 in three sessions. Speech recordings, adaptation data, and model checkpoints remained isolated by DHH speaker. However, correctors’ selections and proposed edits influenced which candidate errors entered the speaker review and recording workflow and could therefore indirectly shape the accepted fine-tuning data. The study was not designed or powered to estimate corrector-level effects separately from speaker-level effects. We treated DHH speakers as the units of ASR analysis ($n=12$) and hearing correctors as the qualitative sample ($n=6$).

\subsection{Procedure}

All sessions were conducted remotely over Zoom and approved by our Institutional Review Board (IRB). Each 90-minute session consisted of four phases: (1) pre-study survey, (2) system tutorial, (3) structured correction-and-adaptation task using EvolveCaptions, and (4) a semi-structured interview. Participants were provided the choice of their preferred accommodation, and all selected Zoom captioning support. Zoom captions were provided only during the tutorial and post-task interview to support communication between participants and researchers, while they were not used during the structured EvolveCaptions tasks.

\subsubsection{Setup and Pre-Survey}

Participants began by completing a demographic and background questionnaire, including questions about prior experiences with ASR and captioning technologies. We also conducted a brief interview with DHH participants about their perceived performance of captioning tools and any frustrations or unmet needs.

\subsubsection{Tutorial and System Orientation}

We introduced participants to EvolveCaptions' collaborative correction and speaker-personalization workflow. DHH users were shown how EvolveCaptions would transcribe their speech and how the hearing user could correct errors. The hearing participants were guided through the correction interface, and the DHH user was introduced to the clause recording interface for between-round personalization.

\subsubsection{Structured Correction-and-Adaptation Task}

Participants completed five scripted reading rounds. In each round, the DHH participant read an approximately 600-word script while EvolveCaptions generated live captions. During the first four rounds, the hearing corrector identified or repaired recognition errors. After each round, the system transformed the selected errors into targeted clauses for the DHH participant to record and then updated the participant's speaker-specific model from the preceding checkpoint. The updated model was used in the next round. Round 5 served as the final post-adaptation measurement and was not followed by another update.

Each participant read the same five scripts once, with their order randomized. We prompted GPT-4 to produce scripts of comparable length and linguistic complexity on everyday topics, including seasonal events, weekend activities, and local news. Audio from the scripted readings was retained for evaluating recognition performance but was excluded from fine-tuning; only the targeted clause recordings were used to update the model. 



\subsubsection{Semi-Structured Interviews}

Following the structured task, we conducted semi-structured interviews lasting \textasciitilde15 minutes with each participant pair. DHH users were asked about perceived caption changes, the recording process, and conditions under which they might use the system. Hearing participants were asked about correction strategies, the effort involved, and their experience of the collaborative workflow. Both participants were also invited to discuss concerns and possible future uses.

All interviews were video recorded and transcribed using Zoom’s ASR, followed by manual correction. Each participant received a \$50 Amazon gift card as compensation.

\subsection{Data Analysis}

\subsubsection{Quantitative Analysis}

We collected system interaction data from all 12 sessions, including: (1) Word Error Rate (WER) for each script pre- and post-adaptation, (2) number of caption corrections made by hearing participants, (3) number and length of clause recordings by DHH participants, and (4) the audio recordings for all five scripted readings completed by each DHH participant, which were later used for the baseline comparisons.

We calculated Word Error Rate (WER) after each iteration, comparing the first and final rounds using the Wilcoxon signed-rank test \cite{robertson2016introduction} to evaluate the statistical significance of improvement. Descriptive statistics ($mean$, $SD$) were used to report correction activity and recording load.

Each DHH participant read all five scripts once, with the script order randomized. To further minimize the chance that the scripts could  differ in recognition difficulty, we conducted a sensitivity analysis testing whether WER decreased across rounds after accounting for both speaker and script identity \cite{zhao2025quantification}. We fitted an ordinary least-squares regression with WER as the outcome and round, coded continuously from 1 to 5, as the predictor. Participant and script identity were included as fixed effects, and standard errors were clustered by participant with a small-sample correction to account for the five repeated measurements from each speaker \cite{colin2015practitioner}.

To contextualize speaker-specific personalization, we also considered three reference baseline conditions:

\begin{itemize}
    \item \textbf{Baseline 1 – Static ASR:} The pretrained Whisper-base model without additional fine-tuning. This model served as the initial checkpoint for each participant and represents general-purpose ASR without speaker-specific adaptation.
    \item \textbf{Baseline 2 – Adapted ASR:} Whisper-base fine-tuned on approximately five minutes of atypical speech (sampled from the TORGO dataset \cite{rudzicz2012torgo}), representing general adaptation without speaker-specific training.
    \item \textbf{Baseline 3 – One-Round Manual Correction:} Captions manually corrected by hearing participants during the first round, serving as a reference for the immediate effect of human transcript repair without model adaptation.
\end{itemize}

Finally, we used descriptive statistics to characterize participants' contributions across the four recording phases. We report the median, range, mean, and standard deviation of the number and total duration of targeted recordings, as well as caption correction activity. Recorded-speech duration measures the amount of audio contributed to personalization; it does not capture the total time or effort involved in reviewing prompts, rerecording speech, correcting captions, or waiting for model updates.

\subsubsection{Qualitative Analysis}
Interview transcripts were analyzed using applied thematic analysis \cite{guest2011applied}. One author reviewed all transcripts and created an initial codebook based on emergent themes using inductive open coding and memoing. This codebook was iteratively refined through discussion with the research team. The final codebook included 4 first-level, 10 second-level, and 25 third-level codes (see Supplementary Material for the full codebook). Two researchers then independently coded the full dataset using the finalized codebook. Inter-rater reliability was calculated using Krippendorff’s alpha \cite{krippendorff2018content} via the ReCal2 package \cite{recal2}. The average $\alpha$ was 0.86 ($\alpha$>0.8 is considered good agreement), with raw agreement of 93\%. Disagreements were resolved via consensus. Codes were then grouped into broader themes and synthesized into a qualitative narrative covering usability, effort, trust, accessibility labor, and system dynamics.

Two authors involved in the analysis—one DHH and one hearing—have prior experience designing accessibility technologies for DHH communities. We acknowledge that these positionalities informed our interpretation of participant narratives, including our attention to both technical performance and lived communication challenges, and that they were critically reflected upon during analysis.

\section{Findings}
We present our findings in two parts. First, the quantitative findings report changes in speaker-specific ASR performance across the five-round protocol and characterize participants' speech and correction contributions. Second, the qualitative findings examine how DHH speakers and hearing correctors experienced their respective roles in the workflow and what considerations shaped their anticipated willingness to use collaborative personalization.

Together, these findings speak to the three design goals introduced in Section~\ref{sec:design}. They characterize the additional contributions involved in personalization, examine recognition changes under a workflow grounded in encountered errors, and show how collaborative correction distributed work between DHH speakers and hearing correctors. The findings also reveal conditions that the design goals alone do not capture: personalization effort was not negligible, improvement was not uniformly monotonic, and collaboration depended on trust, control, and communication context.

\subsection{Quantitative Evaluation of EvolveCaptions}
\label{sec:quant_results}

\begin{figure*}[h]
  \centering
  \includegraphics[
    width=\linewidth,
    alt={Twelve line charts compare word error rate across five rounds for participants P1 through P12. In each chart, the horizontal axis is round and the vertical axis is word error rate, with different vertical scales across participants. The EvolveCaptions line ends lower in Round 5 than in Round 1 for every participant; P1 and P3 show temporary increases in intermediate rounds. Median EvolveCaptions word error rate decreases from 0.495 to 0.345, and Round 5 values range from 0.13 to 0.63. Two fixed-model reference lines show static-base and TORGO-adapted results on the round-specific recordings. A separate marker gives the human-corrected Round 1 value.}
  ]{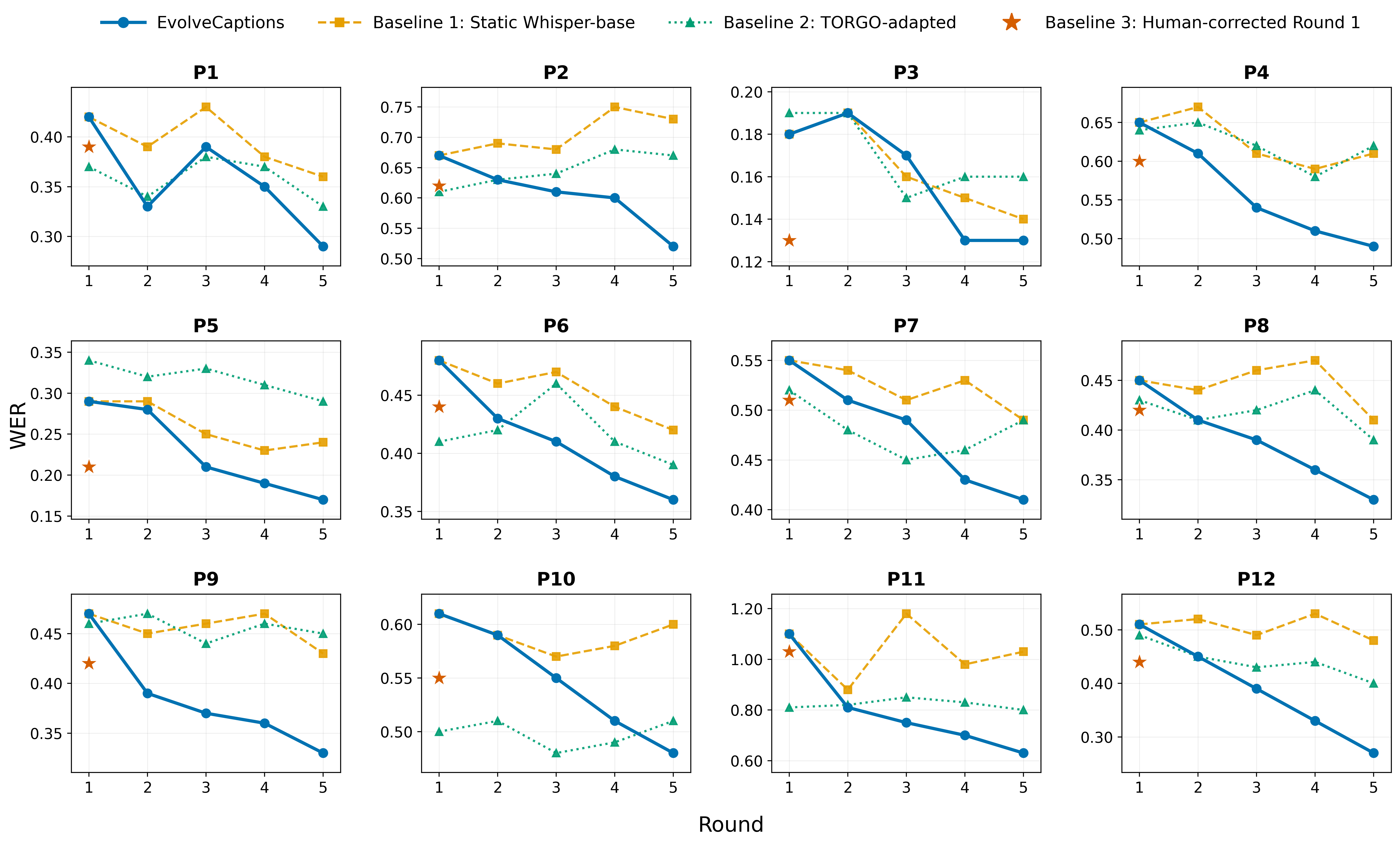}
  \caption{Word Error Rate (WER) improvement across five scripted reading rounds for each DHH participant. EvolveCaptions was updated during the four intervening adaptation phases. Static Baseline 1 and Baseline 2 are fixed-model references evaluated on the same round-specific recordings; their variation across rounds therefore reflects differences in scripts and speech samples rather than model adaptation. The red marker shows the human-corrected transcript from Round 1 (Baseline 3).}
  \Description{Twelve line charts compare word error rate across five rounds for participants P1 through P12. In each chart, the horizontal axis is round and the vertical axis is word error rate, with different vertical scales across participants. The EvolveCaptions line ends lower in Round 5 than in Round 1 for every participant; P1 and P3 show temporary increases in intermediate rounds. Median EvolveCaptions word error rate decreases from 0.495 to 0.345, and Round 5 values range from 0.13 to 0.63. Two fixed-model reference lines show static-base and TORGO-adapted results on the round-specific recordings. A separate marker gives the human-corrected Round 1 value.}
  \label{fig:wer}
\end{figure*}

We first examined how speaker-specific recognition performance changed across the five reading rounds. Across participants, median WER decreased from 0.495 in Round 1 to 0.345 in Round 5. The median within-participant relative reduction was 27.2\% (mean 30.4\%). An exact two-sided Wilcoxon signed-rank test showed that Round 5 WER was significantly lower than Round 1 WER ($W_{+}=78$, $p<.05$; rank-biserial $r_{rb}=1.00$). Figure~\ref{fig:wer} shows the corresponding participant-level trajectories. Although individual trajectories fluctuated across intermediate rounds (e.g., P1 and P3), the overall pattern was downward: all 12 participants had lower WER in Round 5 than in Round 1.

The script-adjusted sensitivity analysis was consistent with the endpoint comparison. After accounting for participant and script identity, WER decreased by an estimated 0.039 per round ($\beta=-0.0394$, 95\% CI [$-0.0583$, $-0.0205$], $p<.05$). Figure~\ref{fig:wer} also compares EvolveCaptions with the static Whisper-base and TORGO-adapted references. By Round 5, EvolveCaptions produced lower WER than both fixed-model references for all 12 participants. Because these reference models did not change across rounds, variation in their WER reflects the different scripts and speech samples. The human-corrected Round 1 transcript shows the immediate effect of correction in that round, rather than improvement retained by the ASR model. Together, the endpoint and script-adjusted results show that speaker-specific WER decreased across the five-round protocol.

\begin{figure*}[h]
  \centering
  \includegraphics[
    width=\linewidth,
    alt={Stacked bars show recording time for 12 participants across four adaptation rounds. The horizontal axis lists P1 through P12, and the vertical axis is recording duration in seconds. Total durations are P1: 302, P2: 228, P3: 173, P4: 535, P5: 289, P6: 263, P7: 268, P8: 271, P9: 221, P10: 278, P11: 457, and P12: 305 seconds. P4 contributes the most, followed by P11; P3 contributes the least. Each bar's four stacked segments divide its total among adaptation Rounds 1 through 4.}
  ]{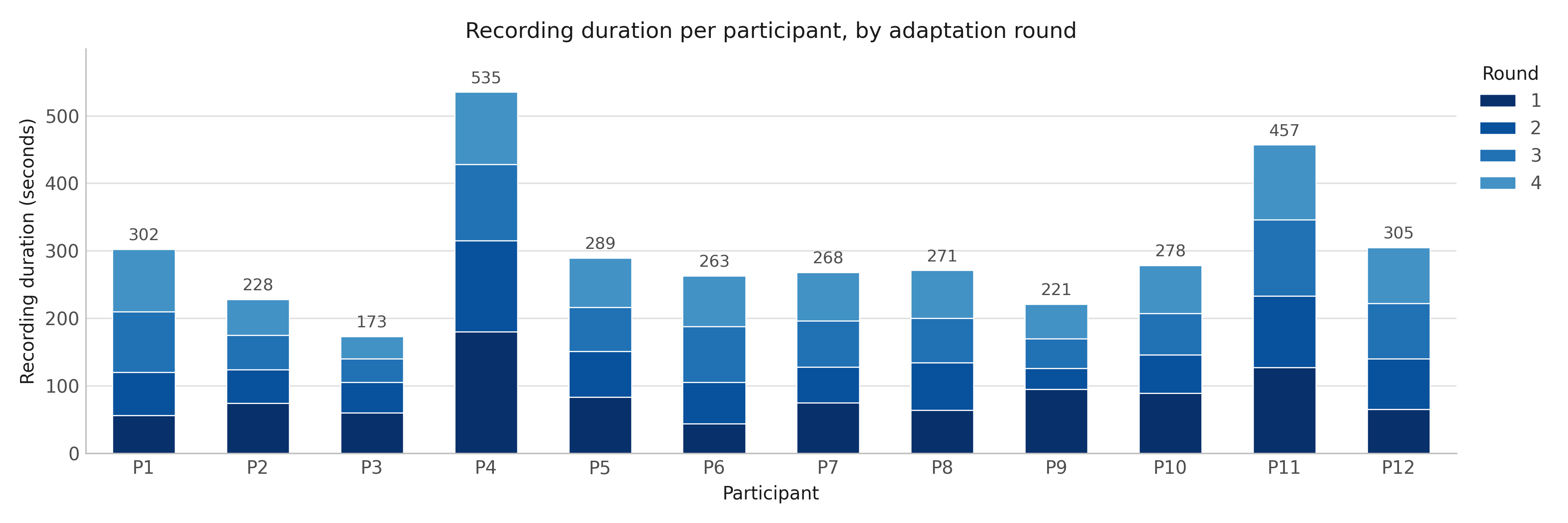}
  \caption{Participant recording durations across four adaptation phases, with segment colors indicating phase-specific durations. Totals are shown above each bar.}
  \Description{Stacked bars show recording time for 12 participants across four adaptation rounds. The horizontal axis lists P1 through P12, and the vertical axis is recording duration in seconds. Total durations are P1: 302, P2: 228, P3: 173, P4: 535, P5: 289, P6: 263, P7: 268, P8: 271, P9: 221, P10: 278, P11: 457, and P12: 305 seconds. P4 contributes the most, followed by P11; P3 contributes the least. Each bar's four stacked segments divide its total among adaptation Rounds 1 through 4.}
  \label{fig:recording}
\end{figure*}

We next quantified the recording, correction, and model-update activities undertaken during personalization. Across the four recording phases, DHH participants contributed a median of 42.5 targeted recordings (range 32--87), totaling a median of 274.5 seconds of speech (range 173--535 seconds; Figure~\ref{fig:recording}). The corresponding means were 46.3 recordings ($SD=14.7$) and 299.2 seconds of speech ($SD=100.6$). Recorded-speech duration does not include time spent reviewing prompts, rerecording, or waiting for model updates.

Across the 12 study sessions, hearing correctors made an average of 72.3 caption edits or uncertainty markings per session ($SD=17.3$). After each recording phase, the system updated the participant's speaker-specific model in under one minute on our NVIDIA T4 deployment. These measures quantify the activities performed by each role; the following qualitative findings examine how DHH speakers and hearing correctors experienced those roles.

\subsection{Qualitative Findings from DHH Participants}

Our qualitative analysis examined how DHH participants experienced their role in the structured correction, recording, and adaptation workflow, as well as what shaped their anticipated willingness to use collaborative personalization in future communication. Through this cross-code sythesis, three themes emerged: (1) participants contrasted the system with existing tools that require them to adapt their speech; (2) they treated recording and correction as strategic, effortful contributions; and (3) they emphasized that such collaboration depends on trust, control, and integration into existing communication practices.

\subsubsection{Adapting the System Rather Than Adapting the Self}

Existing captioning tools were experienced as useful yet persistently misaligned with participants’ speech. Although many participants regularly relied on captions in daily communication, they consistently emphasized that current static ASR systems often failed to accurately capture what they intended to say. These failures were not framed as isolated technical errors, but as part of an ongoing communicative burden. Several participants described captions as altering or distorting meaning. As P7 explained, \textit{``[it] frame \ldots it in a different way,''} while P10 remarked, \textit{``Um, sometimes I do doubt if that's my voice.''} Others noted that captions could substitute contextually plausible words that did not reflect their actual speech. P5 observed, \textit{``It will recognize the context. Not always, but sometimes it will change \ldots the word \ldots because it doesn't fit the context.''}

In response, participants described adapting their own communicative practices to accommodate these limitations. Rather than expecting systems to recognize their speech as-is, many reported slowing down, over-articulating, rehearsing speech, moving to quieter environments, switching platforms, or abandoning speech in favor of typing. These adaptations were necessary for communication to proceed, but were rarely desirable. For some, they also carried emotional costs. P1 described captioning failures during work meetings as anxiety-inducing: \textit{``I get paranoid about, oh God, am I not pronouncing clear enough \ldots and the anxiety starts to kick in.''} Participants also described cases where recognition failed entirely despite effort. As P1 explained, \textit{``There were a couple words where no matter how hard I tried \ldots it just couldn't recognize.''}

These accounts suggest that current captioning systems often preserve access by requiring DHH speakers to adapt themselves to technology. Within this context, EvolveCaptions was found to be compelling because it appeared to invert this relationship. Instead of requiring continuous self-adjustment, participants experienced the system as adapting to them over time.

Crucially, what participants valued was not only improved accuracy, but the visibility of system learning. Several reported noticing that previously misrecognized words were correctly transcribed in later sessions, making adaptation legible and meaningful. P6 reflected, \textit{``I feel that \ldots [the] system [has] been able to get used to my \ldots my speech,''} while P9 described the experience as \textit{``recogniz[ing] my speech patterns \ldots adapt[ing] to how I talk \ldots I feel like I'm teaching the application my voice.''} These responses indicate a shift from passive consumption of imperfect captions to participatory adaptation, where users perceived themselves as actively contributing to a system that increasingly recognized their speech patterns.

This shift also helps explain participants’ willingness to engage with the system despite its added demands. As we show next, however, this willingness was not unconditional, but depended on how participants evaluated the effort required to contribute to system adaptation.

\subsubsection{Recording and Adaptation as Strategic Accessibility Labor}

Recording and adaptation were not experienced as effortless interactions, but as forms of additional work that participants evaluated in relation to their potential benefits. Rather than eliminating accessibility labor, EvolveCaptions redistributed it into a form participants perceived as more purposeful and potentially cumulative. Several described recording as a task, albeit one they were willing to undertake if it led to meaningful improvement. As P2 noted, recording was \textit{``kind of like a \ldots more of a task,''} and later added, \textit{``it's kind of \ldots tiring.''} At the same time, this effort was not rejected outright. Instead, participants weighed it against the value of improved, personalized captions.

This evaluation was strongly context-dependent. When considering future use, participants were generally willing to invest effort in high-stakes scenarios such as work meetings or technical discussions, where misrecognition had greater consequences. P9 explained, \textit{``I'll spend time recording in situations like work meetings \ldots where there's a lot of technical language that I need to catch.''} In contrast, they were less inclined to record in casual or low-stakes interactions, particularly when environmental noise or conversational flow made the effort seem less worthwhile. Personalization was therefore not treated as universally necessary, but as something to be strategically deployed where it mattered most.

Participants also expected this effort to decrease over time. Many described an initial phase of frequent recording, followed by more occasional and targeted contributions once the system had sufficiently adapted. P8 noted a desire to record often at first in order to observe improvement, while P9 explained, \textit{``If I'm to start, I'll say maybe a few times a week \ldots But on a later stage it'll be kind of occasional.''} This expectation highlights that participants were willing to invest effort when they believed it would accumulate toward reduced future burden.

Control over the recording process further shaped its acceptability. Features such as skipping prompts, re-recording, and visual waveform feedback allowed participants to manage when and how they contributed data, making the process feel more flexible and less intrusive. Still, recording was not framed as inherently desirable, but as a strategic investment in future accessibility. As P9 summarized, \textit{``The trade-off is actually worth it \ldots because I know the benefit is [a] more personalized caption system, just for me.''}

Taken together, these accounts suggest that collaborative ASR adaptation should not be characterized as low-friction. Instead, participants treated recording as purposeful accessibility labor whose acceptability depended on whether the effort was bounded, contextually justified, and likely to reduce future demands. At the same time, this willingness to contribute effort was not purely individual, but shaped by the social and relational conditions under which collaboration with a hearing partner occurred.

\subsubsection{Collaboration Requires Trust, Control, and Integration into Existing Workflows}

While participants generally supported collaborative correction, they did not treat all collaborators as equally appropriate. Instead, willingness to involve others depended on trust, familiarity, and concerns about burden. Many preferred close contacts---family members, partners, or long-term colleagues---who were already familiar with their speech patterns. P11 explained they would rely on someone who had known them \textit{``for maybe 5 years or 10 years,''} rather than someone new. Others expressed reluctance to impose on less familiar partners. As P2 stated, \textit{``I don't want to burden colleagues or friends \ldots''} These responses indicate that collaborative correction is not merely a technical feature, but a relational practice shaped by social context.

Participants also raised concerns about how correction might affect the representation of what they said. Correction was not only about improving intelligibility, but about preserving speaker intent. Some worried that incorrect or overly interpretive corrections could distort their meaning. This reveals a central tension in collaborative accessibility: while shared correction can improve recognition, it also introduces dependencies on others and risks around misrepresentation. Notably, one participant suggested that other DHH users might be particularly appropriate collaborators, framing correction as mutual support grounded in shared experience. P3 reflected, \textit{``We are going to benefit from this. We need to do this for each other.''} At the same time, another participant preferred professional captioners, suggesting that trust could also be grounded in expertise rather than familiarity.

Concerns about control extended to the handling of recorded voice data. Participants emphasized that acceptability depended on strong privacy protections and meaningful ownership. Their concerns included voice cloning, unauthorized reuse, and lack of accountability in AI systems. Participants mentioned secure storage, deletion rights, personal ownership, and offline processing as desirable protections. P9 noted, \textit{``The platform should be kind of safe and confidential \ldots It should give me option to delete recordings,''} while P2 expressed concern that others might \textit{``copy my voice as an AI.''} These responses highlight that collaborative personalization introduces not only opportunities for improved access but also heightened stakes around biometric data and identity.

Finally, participants emphasized the importance of integration into existing communication infrastructures. Rather than viewing personalization as a standalone tool, they expected it to be embedded within platforms they already use, such as Zoom or mobile devices. P3 explained, \textit{``I'm hoping that this speech model \ldots be incorporated into \ldots Zoom,''} while others described the need for portability across devices such as phones or smart glasses. Participants also requested customization features such as specialized vocabularies, user accounts, and progress tracking. These expectations suggest that personalization must function as part of a broader ecosystem of communication tools, rather than as an isolated system.

Together, these findings show that collaborative ASR adaptation is not purely a technical improvement, but a socio-technical arrangement. Its success depends on whether users can trust collaborators, maintain control over their data and representation, and access the system within the infrastructures that already shape their communication practices.

\paragraph{Summary of Findings from DHH Participants.}
Across these findings, participants described EvolveCaptions as shifting how accessibility work is experienced during communication. Rather than continuously adapting their speech to static systems, participants engaged with a process in which the system could learn from them over time. However, this shift was conditional. Participants were willing to contribute recording and correction when the effort was justified by perceived long-term benefit, and when collaboration remained grounded in trust, user control, and integration with existing communication tools.

\subsection{Hearing Participants' Experiences}

Findings from hearing participants supported and extended the patterns identified above, particularly around collaboration, effort, and perceived system learning.

All six hearing participants found EvolveCaptions easy to use and responsive, enabling real-time caption correction. While correction introduced a modest cognitive burden, it was generally perceived as manageable. Importantly, participants did not treat this effort as neutral, but evaluated it in relation to its purpose. As H6 noted, \textit{“I would be concerned that [making corrections] will be distracting me \ldots But if this focus can help the system recognize my friend's voice better, I am definitely willing to do it.”} This reflects a broader pattern in which hearing participants understood correction as a meaningful contribution to improving accessibility, rather than as a purely mechanical task.

Similar to DHH participants, willingness to contribute effort was context-dependent. Participants reported being more likely to engage in correction during high-stakes scenarios such as meetings or lectures, where accuracy was critical, and less likely to do so in casual interactions where maintaining conversational flow took priority. They also emphasized relational factors: four participants preferred correcting captions for people they knew well, while two indicated they would assist anyone if asked. 

Participants also noticed and commented on the system’s adaptive behavior. Observing improvements over time reinforced the perception that their efforts contributed to meaningful system learning. As H3 remarked, \textit{“I was impressed by the system adapting to different speakers [who] have so different styles.”} This visibility of adaptation appeared to motivate continued participation, aligning with DHH participants’ descriptions of “teaching” the system.

Hearing participants described strategies to manage cognitive load more effectively. Rather than attempting to correct everything, they selectively prioritized certain types of errors, such as repeated mistakes or content-bearing words. H1 described, \textit{“I tried to follow so I only highlighted uncertain words at first, and then went back to make corrections,”} while H5 noted, \textit{“I focus on repeated words; if I find an error that happens a few times, I will correct them.”} These strategies further suggest that collaborative correction is not a uniform activity, but a learned practice in which participants actively balance attention, effort, and task demands. 

In summary, hearing participants described correction as an attention-demanding but potentially worthwhile contribution, with their anticipated willingness shaped by the stakes of the interaction, their relationship with the DHH speaker, and whether their corrections produced visible improvements.

\section{Discussion}
Our findings build on emerging accessibility frameworks that view communication not as an individual burden but as a shared, relational process. Our qualitative results show that participants evaluated EvolveCaptions not only in terms of accuracy improvements, but in relation to a broader history of adapting themselves to existing captioning systems. In this context, real-time, adaptive captioning was meaningful because it shifted accessibility from individual adaptation toward collaborative, cumulative system learning. In this section, we situate these results in prior work, reflect on user agency and inclusivity in ASR design, and discuss study limitations and implications for future systems.

\subsection{Advancing Communication Accessibility}

\begin{figure}[h]
  \centering
  \includegraphics[
    width=0.6\linewidth,
    alt={Two-by-two diagram classifying captioning systems by adaptation and interaction style. The vertical axis ranges from static to adaptive automatic speech recognition, and the horizontal axis ranges from unilateral to collaborative interaction. Traditional captions occupy the static and unilateral quadrant; collaborative-correction systems occupy static and collaborative; personalized systems such as Project Relate and Project Euphonia occupy adaptive and unilateral. EvolveCaptions alone occupies adaptive and collaborative.}
  ]{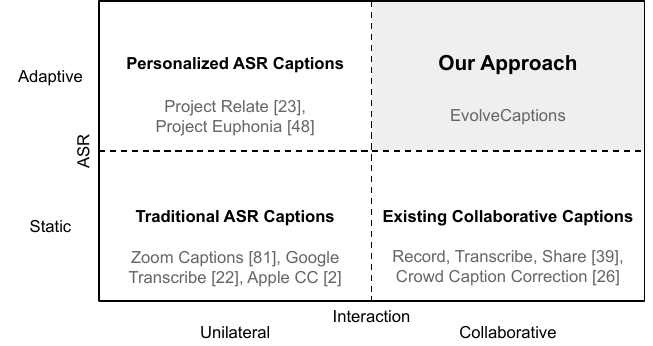}
  \caption{Comparison of captioning technologies for DHH users. The horizontal axis contrasts unilateral versus collaborative interaction, while the vertical axis contrasts static versus adaptive ASR models. EvolveCaptions is positioned in the top-right quadrant, representing a collaborative and adaptive approach.}
  \Description{Two-by-two diagram classifying captioning systems by adaptation and interaction style. The vertical axis ranges from static to adaptive automatic speech recognition, and the horizontal axis ranges from unilateral to collaborative interaction. Traditional captions occupy the static and unilateral quadrant; collaborative-correction systems occupy static and collaborative; personalized systems such as Project Relate and Project Euphonia occupy adaptive and unilateral. EvolveCaptions alone occupies adaptive and collaborative.}
  \label{fig:compare}
\end{figure}

Prior captioning systems for DHH users have primarily relied on static ASR models, which perform poorly on the diverse and often atypical speech patterns of DHH speakers. 

Some systems support speaker adaptation through pre-collected data \cite{martin2025project}, while others offer mechanisms for real-time correction \cite{harrington2013crowd,seita2022remotely}. However, these approaches typically fall short in three key ways: (1) static models do not account for individual variability (Figure \ref{fig:compare}, bottom left); (2) adaptation often places the burden on the DHH speaker (Figure \ref{fig:compare}, top left); and (3) corrections are transient and not retained (Figure \ref{fig:compare}, bottom right). Our qualitative findings show how these limitations are experienced in practice. Participants described existing captioning systems as requiring them to slow down, rehearse speech, or change communication strategies, reinforcing a model in which accessibility is achieved through user adaptation. 

EvolveCaptions addresses these limitations by combining speaker-specific adaptation, real-time collaborative correction, and persistent improvement over time into a unified workflow. By positioning captioning as an evolving socio-technical practice rather than a one-way assistive feature, EvolveCaptions aligns with perspectives on \textit{collaborative accessibility} \cite{branham2015collaborative}, where access is co-produced among participants rather than delivered by a system alone. In our case, this collaboration is centered on the relationship between DHH speakers and hearing partners: hearing partners can correct captions in real time, and these corrections are later integrated into fine-tuning. This design distributes personalization work across both parties while enabling dynamic improvement.

This shift also resonates with Bennett et al.'s framing of \textit{interdependence} in assistive technology \cite{bennett2018interdependence}, which emphasizes that accessible participation is often collective rather than solitary. Our results—particularly participants’ willingness to help and DHH users’ descriptions of “teaching” the system through collaborative correction and recording—show how interdependence can be intentionally designed into adaptive ASR workflows. Rather than eliminating accessibility labor, EvolveCaptions redistributes it across communication partners in ways that participants perceived as more equitable and meaningful.

\subsection{Ability-Based Personalization}

Wobbrock et al. \cite{wobbrock2011ability} propose that systems should adapt to users’ abilities rather than requiring users to adapt to systems. Our findings show that current captioning tools often invert this principle. Participants described actively modifying their speech—slowing down, over-articulating, or avoiding certain words—to accommodate ASR limitations.

EvolveCaptions moves toward ability-based personalization by treating user variability as input to be learned rather than error to be corrected. Participants emphasized that the system allowed them to speak more naturally over time, reducing the need for compensatory behaviors. This reflects a shift from normalization toward accommodation, where the system adapts to diverse speech patterns rather than enforcing conformity.

At the same time, our findings complicate a purely technical interpretation of ability-based design. Personalization was not experienced as effortless or automatic. Instead, participants treated recording and correction as strategic contributions, investing effort when they perceived long-term benefits. This suggests that ability-based systems must account not only for user capabilities, but also for how users evaluate and engage with the work required to support adaptation.

\subsection{Agency and Human-Centered Adaptation}

EvolveCaptions makes adaptation visible and participatory. Rather than operating as a black box, the system exposes a feedback loop in which user contributions directly influence future performance. This aligns with human-centered machine learning principles emphasizing transparency, controllability, and feedback \cite{amershi2019guidelines}.

Participants’ descriptions of “training” or “teaching” the system reflect a shift from passive use toward active participation. Observing improvements over time reinforced a sense of agency, as users could see the effects of their contributions. This contrasts with traditional ASR systems, where adaptation is opaque or absent.

However, our findings also show that agency is not simply a function of visibility. It depends on whether users perceive their contributions as meaningful and cumulative. Participants were willing to engage when they believed that effort would lead to lasting improvement, but less willing when effort seemed repetitive or transient. Designing for agency therefore requires not only making adaptation visible, but ensuring that it is perceived as effective and worthwhile.

\subsection{Social Dynamics of Co-Captioning}

Our findings highlight that collaborative ASR adaptation is inherently social. Participants’ willingness to engage in correction depended on trust, familiarity, and social context. Many preferred close collaborators who understood their speech patterns, while expressing reluctance to burden unfamiliar partners. These dynamics suggest that co-captioning is not a neutral interaction, but one shaped by existing relationships and expectations.

At the same time, collaboration introduces a key tension. While shared correction can improve intelligibility, it also raises concerns about representation and authorship. Participants noted that incorrect corrections could distort what they intended to say, highlighting the risk that collaboration may inadvertently override user voice. This tension underscores that collaborative accessibility must balance shared responsibility with individual control.

These findings extend prior work on collaborative accessibility by showing that participation in accessibility labor is conditional rather than universal. Collaboration is more likely to be accepted when it is socially appropriate and perceived as beneficial for all involved. Designing co-captioning systems therefore requires attention not only to usability, but also to social dynamics and norms of participation.

\subsection{Generalizability to Other Non-Normative Speech}

Although our study focuses on DHH speakers, the underlying framework of EvolveCaptions may generalize to other populations with non-normative speech, including individuals with dysarthria, stroke survivors, non-native speakers, and children. These groups similarly experience mismatches between their speech and models trained on normative speech corpora.

Our findings suggest that iterative and collaborative adaptation may provide an alternative to traditional personalization approaches. By leveraging minimal data and integrating adaptation into everyday interaction, such systems could extend accessibility to a broader range of users without requiring extensive upfront training.

\subsection{Ethical and Deployment Considerations}

Participants’ responses highlight that collaborative personalization introduces new ethical and deployment challenges. Concerns about voice ownership, privacy, and misuse were central to their evaluation of the system. Participants emphasized the importance of deletion rights, secure storage, and, in some cases, offline processing.

Beyond data protection, collaborative correction raises questions about control over representation. Who determines what a user “said” when multiple actors can modify captions? How should consent be managed in shared interactions? These questions point to the need for governance mechanisms that account for both individual and collective contributions to model adaptation.

Participants also emphasized the importance of integration into existing communication infrastructures. Rather than adopting standalone tools, they expected personalization to be embedded within platforms such as Zoom or mobile devices. This suggests that deployment success depends not only on technical capability, but on alignment with existing ecosystems and workflows.

\subsection{Limitations and Future Work}

While EvolveCaptions demonstrates the potential of collaborative, between-round ASR adaptation, our findings also highlight several limitations and open challenges for deploying such systems in practice. Importantly, several of these limitations are not only technical but socio-technical, reflecting the effort, trust, and coordination required to sustain collaborative accessibility in real-world settings.

\textbf{Controlled Evaluation Context.} We conducted a lab-style, remote study with scripted speech and dyadic interaction. This controlled setup enabled us to isolate adaptation effects and analyze behavioral patterns systematically. However, scripted reading does not fully capture the dynamics of spontaneous conversation, such as interruptions, overlapping speech, or contextual disfluencies. Future work will involve longitudinal field deployments of EvolveCaptions in real-world communication settings, such as classrooms, workplaces, and video conferencing platforms (e.g., Zoom, Google Meet), to evaluate the system's adaptability under more natural conditions and interactional diversity.

\textbf{System Deployment Constraints.} For this evaluation, EvolveCaptions was deployed on a cloud-based server to ensure consistency across participants. While the underlying pipeline can run locally, the cloud setup allowed us to control for hardware variability and focus on testing adaptation mechanisms. To support real-world deployment, future work will explore lightweight on-device fine-tuning, real-time model updates under limited compute, and robustness in non-ideal conditions such as noisy environments or poor microphone quality.

\textbf{Collaborative Correction and Cognitive Load.} Although hearing participants reported that real-time correction was manageable and meaningful, the cognitive effort required---especially during fast-paced dialogue---remains a challenge. In long-term use or high-pressure scenarios, this effort may become unsustainable. Future work should explore interaction designs that reduce correction load while maintaining effectiveness, as well as evaluate hearing participants’ experience more rigorously using tools such as NASA-TLX \cite{cao2009nasa} or post-task comprehension assessments.

\textbf{Long-Term Personalization.} While EvolveCaptions supports incremental adaptation, it does not yet account for temporal variability in speech, such as changes due to fatigue, illness, stress, or post-surgical recovery. Future systems must incorporate mechanisms to detect when personalization becomes outdated, offer options for re-adaptation, and allow users to manage and synchronize their personalized models across devices and contexts. Supporting longitudinal flexibility will be critical for ensuring that adaptive captioning remains robust and equitable over time.

\textbf{Interaction Dynamics.} Previous work has explored behavioral patterns in ASR-mediated communication, such as hearing speakers slowing down or DHH participants modifying gaze behavior \cite{seita2018behavioral, seita2020deaf, seita2021deaf}. Our study adds a new dimension to this interaction: hearing participants actively monitored and corrected captions in real time, while DHH participants selectively recorded clauses based on those corrections. These findings reinforce that ASR personalization is not only technical but socio-technical, shaped by trust, familiarity, and the distribution of accessibility labor. Future work could systematically examine these co-adaptive behaviors across diverse speaker dyads to better understand how interpersonal dynamics influence long-term adoption and effectiveness.

\textbf{Carryover Learning Effects.} Four of the six hearing correctors participated in sessions with multiple DHH speakers. Although correctors did not contribute speech recordings, their selections, proposed edits, and uncertainty markings influenced which candidate errors entered the DHH review and recording workflow. Experience across sessions could therefore have changed the number or composition of the training examples ultimately accepted by DHH speakers and, indirectly, the resulting model trajectories. Because the study was not designed to separate corrector experience from speaker differences, we cannot estimate an independent corrector-learning effect. We therefore interpret the accuracy results as outcomes of the complete collaborative workflow. Future studies should counterbalance corrector experience or explicitly model corrector identity and session order in a larger sample.

\section{Conclusion}
EvolveCaptions demonstrates a new approach to ASR personalization for DHH speakers by combining live human correction, targeted data collection, and collaborative interaction. By involving hearing participants in the correction process and focusing training on misrecognized speech, the system enables real-time, incremental adaptation grounded in principles of collective access. Our evaluation with 12 DHH and six hearing participants showed that even a small set of targeted recordings can substantially improve captioning quality. Beyond these improvements, our findings highlight how such systems can shift accessibility from requiring DHH users to adapt to technology toward collaborative, cumulative adaptation. At the same time, participants engaged with recording and correction when the effort was perceived as worthwhile and when collaboration was grounded in trust and control. Together, these results point toward ASR systems that improve accuracy while redistributing accessibility work in more equitable and sustainable ways.

\bibliographystyle{ACM-Reference-Format}
\bibliography{ref}

@article{huang2023contextualized,
  title={Contextualized end-to-end speech recognition with contextual phrase prediction network},
  author={Huang, Kaixun and Zhang, Ao and Yang, Zhanheng and Guo, Pengcheng and Mu, Bingshen and Xu, Tianyi and Xie, Lei},
  journal={arXiv preprint arXiv:2305.12493},
  year={2023}
}

@inproceedings{wang2024can,
  title={Can whisper perform speech-based in-context learning?},
  author={Wang, Siyin and Yang, Chao-Han and Wu, Ji and Zhang, Chao},
  booktitle={ICASSP 2024-2024 IEEE International Conference on Acoustics, Speech and Signal Processing (ICASSP)},
  pages={13421--13425},
  year={2024},
  organization={IEEE}
}

@inproceedings{pokel2026demonstration,
  title={Demonstration of Adapt4Me: An Uncertainty-Aware Authoring Environment for Personalizing Automatic Speech Recognition to Non-normative Speech},
  author={Pokel, Niclas and Zhao, Yiming and Moure, Pehuen and Gao, Yingqiang and Boehringer, Roman},
  booktitle={Proceedings of the 8th ACM Conference on Conversational User Interfaces},
  pages={1--7},
  year={2026}
}

@inproceedings{kuhn2025communication,
  title={Communication access real-time translation through collaborative correction of automatic speech recognition},
  author={Kuhn, Korbinian and Kersken, Verena and Zimmermann, Gottfried},
  booktitle={Proceedings of the Extended Abstracts of the CHI Conference on Human Factors in Computing Systems},
  pages={1--8},
  year={2025}
}

@inproceedings{shah2026claris,
  title={CLARIS: Clear and Intelligible Speech from Whispered and Dysarthric Voices},
  author={Shah, Neil and Sonkar, Yash and Subhash Karande, Shirish and Gandhi, Vineet},
  booktitle={Proceedings of the 2026 CHI Conference on Human Factors in Computing Systems},
  pages={1--18},
  year={2026}
}

@article{huber2026adapting,
  title={Adapting Foundation ASR Models to Dysarthric Speech: A Case Study},
  author={Huber, Christian and Kernahan, Laura and Waibel, Alexander},
  journal={arXiv preprint arXiv:2606.31722},
  year={2026}
}

@inproceedings{hernandez2026confidence,
  title={Confidence-guided error correction for disordered speech recognition},
  author={Hernandez, Abner and Arias-Vergara, Tom{\'a}s and Maier, Andreas and P{\'e}rez-Toro, Paula Andrea},
  booktitle={ICASSP 2026-2026 IEEE International Conference on Acoustics, Speech and Signal Processing (ICASSP)},
  pages={18447--18451},
  year={2026},
  organization={IEEE}
}

@article{colin2015practitioner,
  title={A practitioner’s guide to cluster-robust inference},
  author={Colin Cameron, A and Miller, Douglas L},
  journal={Journal of human resources},
  volume={50},
  number={2},
  pages={317--372},
  year={2015},
  publisher={University of Wisconsin Press}
}

@inproceedings{wu2025soundnarratives,
  title={SoundNarratives: Rich Auditory Scene Descriptions to Support Deaf and Hard of Hearing People},
  author={Wu, Liang-Yuan and Jain, Dhruv},
  booktitle={Proceedings of the 27th International ACM SIGACCESS Conference on Computers and Accessibility},
  pages={1--15},
  year={2025}
}

@inproceedings{wu2025cartgpt,
  title={CARTGPT: Real-Time Correction of CART Captions Using Large Language Models},
  author={Wu, Liang-Yuan and Kleiver, Andrea and Jain, Dhruv},
  booktitle={Proceedings of the 27th International ACM SIGACCESS Conference on Computers and Accessibility},
  pages={1--11},
  year={2025}
}

@article{hughes2026assessing,
  title={Assessing the Role of Medical Caption Technology to Support Physician-Patient Communication for Patients With Hearing Loss: Mixed Methods Pilot Study},
  author={Hughes, Sarah E and Wu, Liang-Yuan and Ma, Lindsay J and Jain, Dhruv and McKee, Michael M},
  journal={JMIR Rehabilitation and Assistive Technologies},
  volume={13},
  number={1},
  pages={e79073},
  year={2026},
  publisher={JMIR Publications Inc., Toronto, Canada}
}

@misc{cartoverview,
  author={UW DO-IT},
  title={What is real-time captioning?},
  year={2021},
  url={https://www.washington.edu/doit/what-real-time-captioning},
  note={Accessed: 2025-03-10}
}

@incollection{cavender2008hearing,
  title={Hearing impairments},
  author={Cavender, Anna and Ladner, Richard E},
  booktitle={Web accessibility: A foundation for research},
  pages={25--35},
  year={2008},
  publisher={Springer}
}

@book{ladd2003understanding,
  title={Understanding deaf culture: In search of deafhood},
  author={Ladd, Paddy},
  year={2003},
  publisher={Multilingual Matters}
}

@inproceedings{loizides2020breaking,
  title={Breaking boundaries with live transcribe: Expanding use cases beyond standard captioning scenarios},
  author={Loizides, Fernando and Basson, Sara and Kanevsky, Dimitri and Prilepova, Olga and Savla, Sagar and Zaraysky, Susanna},
  booktitle={Proceedings of the 22nd International ACM SIGACCESS Conference on Computers and Accessibility},
  pages={1--6},
  year={2020}
}

@inproceedings{kawas2016improving,
  title={Improving real-time captioning experiences for deaf and hard of hearing students},
  author={Kawas, Saba and Karalis, George and Wen, Tzu and Ladner, Richard E},
  booktitle={Proceedings of the 18th International ACM SIGACCESS Conference on Computers and Accessibility},
  pages={15--23},
  year={2016}
}

@article{kushalnagar2014accessibility,
  title={Accessibility evaluation of classroom captions},
  author={Kushalnagar, Raja S and Lasecki, Walter S and Bigham, Jeffrey P},
  journal={ACM Transactions on Accessible Computing (TACCESS)},
  volume={5},
  number={3},
  pages={1--24},
  year={2014},
  publisher={ACM New York, NY, USA}
}

@inproceedings{cerva2012real,
  title={Real-Time Lecture Transcription using ASR for Czech Hearing Impaired or Deaf Students.},
  author={Cerva, Petr and Silovsk{\`y}, Jan and Zd{\'a}nsk{\`y}, Jindrich and Nouza, Jan and Malek, Jiri},
  booktitle={INTERSPEECH},
  pages={763--766},
  year={2012}
}

@inproceedings{elliot2016deaf,
  title={Deaf and hard of hearing individuals' perceptions of communication with hearing colleagues in small groups},
  author={Elliot, Lisa and Stinson, Michael and Mallory, James and Easton, Donna and Huenerfauth, Matt},
  booktitle={Proceedings of the 18th International ACM SIGACCESS Conference on Computers and Accessibility},
  pages={271--272},
  year={2016}
}

@inproceedings{prietch2014speech,
  title={A speech-to-text system’s acceptance evaluation: would deaf individuals adopt this technology in their lives?},
  author={Prietch, Soraia Silva and de Souza, Napoliana Silva and Filgueiras, Lucia Villela Leite},
  booktitle={International Conference on Universal Access in Human-Computer Interaction},
  pages={440--449},
  year={2014},
  organization={Springer}
}

@inproceedings{lasecki2012real,
  title={Real-time captioning by groups of non-experts},
  author={Lasecki, Walter and Miller, Christopher and Sadilek, Adam and Abumoussa, Andrew and Borrello, Donato and Kushalnagar, Raja and Bigham, Jeffrey},
  booktitle={Proceedings of the 25th annual ACM symposium on User interface software and technology},
  pages={23--34},
  year={2012}
}

@article{wald2008universal,
  title={Universal access to communication and learning: the role of automatic speech recognition},
  author={Wald, Mike and Bain, Keith},
  journal={Universal Access in the Information Society},
  volume={6},
  number={4},
  pages={435--447},
  year={2008},
  publisher={Springer}
}

@inproceedings{elliot2017user,
  title={User experiences when testing a messaging app for communication between individuals who are hearing and deaf or hard of hearing},
  author={Elliot, Lisa B and Stinson, Michael and Ahmed, Syed and Easton, Donna},
  booktitle={Proceedings of the 19th International ACM SIGACCESS Conference on Computers and Accessibility},
  pages={405--406},
  year={2017}
}

@inproceedings{seita2022remotely,
  title={Remotely co-designing features for communication applications using automatic captioning with deaf and hearing pairs},
  author={Seita, Matthew and Lee, Sooyeon and Andrew, Sarah and Shinohara, Kristen and Huenerfauth, Matt},
  booktitle={Proceedings of the 2022 CHI Conference on Human Factors in Computing Systems},
  pages={1--13},
  year={2022}
}

@inproceedings{yamamoto2021see,
  title={See-through captions: real-time captioning on transparent display for deaf and hard-of-hearing people},
  author={Yamamoto, Kenta and Suzuki, Ippei and Shitara, Akihisa and Ochiai, Yoichi},
  booktitle={Proceedings of the 23rd International ACM SIGACCESS Conference on Computers and Accessibility},
  pages={1--4},
  year={2021}
}

@inproceedings{berke2017deaf,
  title={Deaf and hard-of-hearing perspectives on imperfect automatic speech recognition for captioning one-on-one meetings},
  author={Berke, Larwan and Caulfield, Christopher and Huenerfauth, Matt},
  booktitle={Proceedings of the 19th International ACM SIGACCESS Conference on Computers and Accessibility},
  pages={155--164},
  year={2017}
}

@inproceedings{bigham2017deaf,
  title={On how deaf people might use speech to control devices},
  author={Bigham, Jeffrey P and Kushalnagar, Raja and Huang, Ting-Hao Kenneth and Flores, Juan Pablo and Savage, Saiph},
  booktitle={Proceedings of the 19th international ACM SIGACCESS conference on computers and accessibility},
  pages={383--384},
  year={2017}
}

@book{mattys2013speech,
  title={Speech recognition in adverse conditions: Explorations in behaviour and neuroscience},
  author={Mattys, Sven and Bradlow, Ann and Davis, Matthew and Scott, Sophie},
  year={2013},
  publisher={Psychology Press}
}

@inproceedings{pradhan2018accessibility,
  title={" Accessibility Came by Accident" Use of Voice-Controlled Intelligent Personal Assistants by People with Disabilities},
  author={Pradhan, Alisha and Mehta, Kanika and Findlater, Leah},
  booktitle={Proceedings of the 2018 CHI Conference on human factors in computing systems},
  pages={1--13},
  year={2018}
}

@inproceedings{mengistu2011comparing,
  title={Comparing humans and automatic speech recognition systems in recognizing dysarthric speech},
  author={Mengistu, Kinfe Tadesse and Rudzicz, Frank},
  booktitle={Canadian Conference on Artificial Intelligence},
  pages={291--300},
  year={2011},
  organization={Springer}
}

@article{zhao2025quantification,
  title={Quantification of automatic speech recognition system performance on d/deaf and hard of hearing speech},
  author={Zhao, Robin and Choi, Anna SG and Koenecke, Allison and Rameau, Ana{\"\i}s},
  journal={The Laryngoscope},
  volume={135},
  number={1},
  pages={191--197},
  year={2025},
  publisher={Wiley Online Library}
}

@article{rodolitz2019accessibility,
  title={Accessibility of voice-activated agents for people who are deaf or hard of hearing},
  author={Rodolitz, Jason and Gambill, Evan and Willis, Brittany and Vogler, Christian and Kushalnagar, Raja},
  journal={Journal on Technology and Persons with Disabilities},
  volume={7},
  pages={144--156},
  year={2019}
}

@article{gottermeier2016user,
  title={User evaluation of automatic speech recognition systems for deaf-hearing interactions at school and work},
  author={Gottermeier, Linda G and DE FILIPPO, CAROL L and KUSHALNAGAR, R AJA and BASTIAN, BONNIE L},
  journal={Audiology Today},
  volume={28},
  number={2},
  pages={20--34},
  year={2016}
}

@article{de2010longitudinal,
  title={A longitudinal study on auditory perception and speech intelligibility in deaf children implanted younger than 18 months in comparison to those implanted at later ages},
  author={De Raeve, Leo},
  journal={Otology \& Neurotology},
  volume={31},
  number={8},
  pages={1261--1267},
  year={2010},
  publisher={LWW}
}

@article{baskar2022speaker,
  title={Speaker adaptation for Wav2vec2 based dysarthric ASR},
  author={Baskar, Murali Karthick and Herzig, Tim and Nguyen, Diana and Diez, Mireia and Polzehl, Tim and Burget, Luk{\'a}{\v{s}} and {\v{C}}ernock{\`y}, Jan and others},
  journal={arXiv preprint arXiv:2204.00770},
  year={2022}
}

@article{jiang2024perceiver,
  title={Perceiver-prompt: Flexible speaker adaptation in Whisper for chinese disordered speech recognition},
  author={Jiang, Yicong and Wang, Tianzi and Xie, Xurong and Liu, Juan and Sun, Wei and Yan, Nan and Chen, Hui and Wang, Lan and Liu, Xunying and Tian, Feng},
  journal={arXiv preprint arXiv:2406.09873},
  year={2024}
}

@inproceedings{macdonald2021disordered,
  title={Disordered Speech Data Collection: Lessons Learned at 1 Million Utterances from Project Euphonia.},
  author={MacDonald, Robert L and Jiang, Pan-Pan and Cattiau, Julie and Heywood, Rus and Cave, Richard and Seaver, Katie and Ladewig, Marilyn A and Tobin, Jimmy and Brenner, Michael P and Nelson, Philip C and others},
  booktitle={Interspeech},
  volume={2021},
  pages={4833--4837},
  year={2021}
}

@inproceedings{tobin2022personalized,
  title={Personalized automatic speech recognition trained on small disordered speech datasets},
  author={Tobin, Jimmy and Tomanek, Katrin},
  booktitle={ICASSP 2022-2022 IEEE International Conference on Acoustics, Speech and Signal Processing (ICASSP)},
  pages={6637--6641},
  year={2022},
  organization={IEEE}
}

@inproceedings{harrington2013crowd,
  title={Crowd caption correction (ccc)},
  author={Harrington, Rebecca Perkins and Vanderheiden, Gregg C},
  booktitle={Proceedings of the 15th International ACM SIGACCESS Conference on Computers and Accessibility},
  pages={1--2},
  year={2013}
}

@inproceedings{mcdonnell2022understanding,
  title={Understanding social and environmental factors to enable collective access approaches to the design of captioning technology},
  author={McDonnell, Emma},
  booktitle={Proceedings of the 24th International ACM SIGACCESS Conference on Computers and Accessibility},
  pages={1--8},
  year={2022}
}

@inproceedings{mcdonnell2023easier,
  title={“Easier or Harder, Depending on Who the Hearing Person Is”: Codesigning Videoconferencing Tools for Small Groups with Mixed Hearing Status},
  author={McDonnell, Emma J and Moon, Soo Hyun and Jiang, Lucy and Goodman, Steven M and Kushalnagar, Raja and Froehlich, Jon E and Findlater, Leah},
  booktitle={Proceedings of the 2023 CHI Conference on Human Factors in Computing Systems},
  pages={1--15},
  year={2023}
}

@inproceedings{kuhn2024record,
  title={Record, Transcribe, Share: An Accessible Open-Source Video Platform for Deaf and Hard of Hearing Viewers},
  author={Kuhn, Korbinian and Reuter, Benedikt and Egger, Niklas and Zimmermann, Gottfried},
  booktitle={Proceedings of the 26th International ACM SIGACCESS Conference on Computers and Accessibility},
  pages={1--6},
  year={2024}
}

@inproceedings{fok2018towards,
  title={Towards more robust speech interactions for deaf and hard of hearing users},
  author={Fok, Raymond and Kaur, Harmanpreet and Palani, Skanda and Mott, Martez E and Lasecki, Walter S},
  booktitle={Proceedings of the 20th international ACM SIGACCESS conference on computers and accessibility},
  pages={57--67},
  year={2018}
}

@inproceedings{radford2023robust,
  title={Robust speech recognition via large-scale weak supervision},
  author={Radford, Alec and Kim, Jong Wook and Xu, Tao and Brockman, Greg and McLeavey, Christine and Sutskever, Ilya},
  booktitle={International conference on machine learning},
  pages={28492--28518},
  year={2023},
  organization={PMLR}
}

@article{rowe2022characterizing,
  title={Characterizing dysarthria diversity for automatic speech recognition: A tutorial from the clinical perspective},
  author={Rowe, Hannah P and Gutz, Sarah E and Maffei, Marc F and Tomanek, Katrin and Green, Jordan R},
  journal={Frontiers in computer science},
  volume={4},
  pages={770210},
  year={2022},
  publisher={Frontiers Media SA}
}

@article{ghimire2024comprehensive,
  title={A Comprehensive Study of the Current State-of-the-Art in Nepali Automatic Speech Recognition Systems},
  author={Ghimire, Rupak Raj and Bal, Bal Krishna and Poudyal, Prakash},
  journal={arXiv preprint arXiv:2402.03050},
  year={2024}
}

@inproceedings{glasser2017deaf,
  title={Deaf, hard of hearing, and hearing perspectives on using automatic speech recognition in conversation},
  author={Glasser, Abraham and Kushalnagar, Kesavan and Kushalnagar, Raja},
  booktitle={Proceedings of the 19th International ACM SIGACCESS Conference on Computers and Accessibility},
  pages={427--432},
  year={2017}
}

@article{martin2025project,
  title={Project Euphonia: advancing inclusive speech recognition through expanded data collection and evaluation},
  author={Martin, Alicia and MacDonald, Robert L and Jiang, Pan-Pan and Ladewig, Marilyn and Cattiau, Julie and Heywood, Rus and Cave, Richard and Tobin, Jimmy and Nelson, Philip C and Tomanek, Katrin},
  journal={Frontiers in Language Sciences},
  volume={4},
  pages={1569448},
  year={2025},
  publisher={Frontiers Media SA}
}

@article{iezzoni2004communicating,
  title={Communicating about health care: observations from persons who are deaf or hard of hearing},
  author={Iezzoni, Lisa I and O'Day, Bonnie L and Killeen, Mary and Harker, Heather},
  journal={Annals of internal medicine},
  volume={140},
  number={5},
  pages={356--362},
  year={2004},
  publisher={American College of Physicians}
}

@misc{google_relate,
  author = {{Google Research}},
  title = {Project Relate},
  howpublished = {\url{https://sites.research.google/relate/}},
  year = {2025},
  note = {Accessed: 2025-09-08},
}

@article{millett2021accuracy,
  title={Accuracy of Speech-to-Text Captioning for Students Who are Deaf or Hard of Hearing.},
  author={Millett, Pam},
  journal={Journal of Educational, Pediatric \& (Re) Habilitative Audiology},
  volume={25},
  year={2021}
}

@inproceedings{jain2018exploring,
  title={Exploring augmented reality approaches to real-time captioning: A preliminary autoethnographic study},
  author={Jain, Dhruv and Chinh, Bonnie and Findlater, Leah and Kushalnagar, Raja and Froehlich, Jon},
  booktitle={Proceedings of the 2018 ACM Conference Companion Publication on Designing Interactive Systems},
  pages={7--11},
  year={2018}
}

@article{moore1993hearing,
  title={For Hearing People Only: Answers to some of the most commonly asked questions about the deaf community, its culture, and the" deaf reality"},
  author={Moore, Matthew S and Levitan, Linda},
  journal={(No Title)},
  year={1993}
}

@inproceedings{kafle2017evaluating,
  title={Evaluating the usability of automatically generated captions for people who are deaf or hard of hearing},
  author={Kafle, Sushant and Huenerfauth, Matt},
  booktitle={Proceedings of the 19th International ACM SIGACCESS Conference on Computers and Accessibility},
  pages={165--174},
  year={2017}
}

@inproceedings{shor2019personalizing,
  title     = {Personalizing ASR for Dysarthric and Accented Speech with Limited Data},
  author    = {Shor, Joel and Emanuel, Dotan and Lang, Oran and Tuval, Omry and Brenner, Michael and Cattiau, Julie and others},
  booktitle = {Proceedings of Interspeech 2019},
  year      = {2019},
  pages     = {784--788},
  doi       = {10.21437/Interspeech.2019-1427}
}

@inproceedings{rudzicz2010towards,
  title={Towards a noisy-channel model of dysarthria in speech recognition},
  author={Rudzicz, Frank},
  booktitle={Proceedings of the NAACL HLT 2010 Workshop on Speech and Language Processing for Assistive Technologies},
  pages={80--88},
  year={2010}
}

@article{mulfari2024voice,
  title={A voice user interface on the edge for people with speech impairments},
  author={Mulfari, Davide and Villari, Massimo},
  journal={Electronics},
  volume={13},
  number={7},
  pages={1389},
  year={2024},
  publisher={MDPI}
}

@incollection{robertson2016introduction,
  title={An introduction to modern statistical methods in HCI},
  author={Robertson, Judy and Kaptein, Maurits},
  booktitle={Modern Statistical Methods for HCI},
  pages={1--14},
  year={2016},
  publisher={Springer}
}

@inproceedings{chen2024towards,
  title={Towards Co-Creating Access and Inclusion: A Group Autoethnography on a Hearing Individual's Journey Towards Effective Communication in Mixed-Hearing Ability Higher Education Settings},
  author={Chen, Si and Waller, James and Seita, Matthew and Vogler, Christian and Kushalnagar, Raja and Wang, Qi},
  booktitle={Proceedings of the 2024 CHI Conference on Human Factors in Computing Systems},
  pages={1--14},
  year={2024}
}

@book{guest2011applied,
  title={Applied thematic analysis},
  author={Guest, Greg and MacQueen, Kathleen M and Namey, Emily E},
  year={2011},
  publisher={sage publications}
}

@book{krippendorff2018content,
  title={Content analysis: An introduction to its methodology},
  author={Krippendorff, Klaus},
  year={2018},
  publisher={Sage publications}
}

@misc{recal2,
  author={Deen Freelon},
  title={ReCal2: Reliability for 2 Coders},
  year={2010},
  url={http://dfreelon.org/utils/recalfront/recal2/},
  note={Accessed: 2025-03-10}
}

@misc{apple_cc,
  author={Apple},
  title={Get live captions of spoken audios on iPhone},
  year={2025},
  url={https://support.apple.com/guide/iphone/get-live-captions-of-spoken-audio-iphe0990f7bb/ios},
  note={Accessed: 2025-03-10}
}

@misc{google_transcribe,
  author={Google},
  title={Live Transcribe \& notification},
  year={2025},
  url={https://play.google.com/store/apps/details?id=com.google.audio.hearing.visualization.accessibility.scribe},
  note={Accessed: 2025-03-10}
}

@inproceedings{seita2018behavioral,
  title={Behavioral changes in speakers who are automatically captioned in meetings with deaf or hard-of-hearing peers},
  author={Seita, Matthew and Albusays, Khaled and Kafle, Sushant and Stinson, Michael and Huenerfauth, Matt},
  booktitle={Proceedings of the 20th International ACM SIGACCESS Conference on Computers and Accessibility},
  pages={68--80},
  year={2018}
}

@inproceedings{seita2021deaf,
  title={Deaf and hard-of-hearing users' preferences for hearing speakers' behavior during technology-mediated in-person and remote conversations},
  author={Seita, Matthew and Andrew, Sarah and Huenerfauth, Matt},
  booktitle={Proceedings of the 18th International Web for All Conference},
  pages={1--12},
  year={2021}
}

@inproceedings{seita2020deaf,
  title={Deaf individuals' views on speaking behaviors of hearing peers when using an automatic captioning app},
  author={Seita, Matthew and Huenerfauth, Matt},
  booktitle={Extended Abstracts of the 2020 CHI Conference on Human Factors in Computing Systems},
  pages={1--8},
  year={2020}
}

@inproceedings{kushalnagar2020teleconference,
  title={Teleconference accessibility and guidelines for deaf and hard of hearing users},
  author={Kushalnagar, Raja S and Vogler, Christian},
  booktitle={Proceedings of the 22nd International ACM SIGACCESS Conference on Computers and Accessibility},
  pages={1--6},
  year={2020}
}

@inproceedings{kushalnagar2013captions,
  title={Captions versus transcripts for online video content},
  author={Kushalnagar, Raja S and Lasecki, Walter S and Bigham, Jeffrey P},
  booktitle={Proceedings of the 10th International Cross-Disciplinary Conference on Web Accessibility},
  pages={1--4},
  year={2013}
}

@article{wobbrock2011ability,
  title={Ability-based design: Concept, principles and examples},
  author={Wobbrock, Jacob O and Kane, Shaun K and Gajos, Krzysztof Z and Harada, Susumu and Froehlich, Jon},
  journal={ACM Transactions on Accessible Computing (TACCESS)},
  volume={3},
  number={3},
  pages={1--27},
  year={2011},
  publisher={ACM New York, NY, USA}
}

@inproceedings{amershi2019guidelines,
  title={Guidelines for human-AI interaction},
  author={Amershi, Saleema and Weld, Dan and Vorvoreanu, Mihaela and Fourney, Adam and Nushi, Besmira and Collisson, Penny and Suh, Jina and Iqbal, Shamsi and Bennett, Paul N and Inkpen, Kori and others},
  booktitle={Proceedings of the 2019 chi conference on human factors in computing systems},
  pages={1--13},
  year={2019}
}

@article{rudzicz2012torgo,
  title={The TORGO database of acoustic and articulatory speech from speakers with dysarthria},
  author={Rudzicz, Frank and Namasivayam, Aravind Kumar and Wolff, Talya},
  journal={Language resources and evaluation},
  volume={46},
  number={4},
  pages={523--541},
  year={2012},
  publisher={Springer}
}

@article{cao2009nasa,
  title={NASA TLX: Software for assessing subjective mental workload},
  author={Cao, Alex and Chintamani, Keshav K and Pandya, Abhilash K and Ellis, R Darin},
  journal={Behavior research methods},
  volume={41},
  number={1},
  pages={113--117},
  year={2009},
  publisher={Springer}
}

@misc{reactjs,
  author={Meta Open Source},
  title={React: The library for web and native user interfaces},
  year={2025},
  url={https://react.dev},
  note={Accessed: 2025-09-10}
}

@misc{vite,
  author={VoidZero Inc. \& Vite Contributors},
  title={Vite: Next Generation Frontend Tooling},
  year={2025},
  url={https://vite.dev},
  note={Accessed: 2025-09-10}
}

@misc{audioworklet,
  author={Mozilla},
  title={AudioWorklet - Web APIs},
  year={2025},
  url={https://developer.mozilla.org/en-US/docs/Web/API/AudioWorklet},
  note={Accessed: 2025-09-10}
}

@inproceedings{branham2015collaborative,
  title={Collaborative accessibility: How blind and sighted companions co-create accessible home spaces},
  author={Branham, Stacy M and Kane, Shaun K},
  booktitle={Proceedings of the 33rd Annual ACM Conference on Human Factors in Computing Systems},
  pages={2373--2382},
  year={2015}
}

@inproceedings{bennett2018interdependence,
  title={Interdependence as a frame for assistive technology research and design},
  author={Bennett, Cynthia L and Brady, Erin and Branham, Stacy M},
  booktitle={Proceedings of the 20th international acm sigaccess conference on computers and accessibility},
  pages={161--173},
  year={2018}
}

@misc{zoom_captions,
  author={Zoom},
  title={Viewing captions in a meeting or webinar},
  year={2025},
  url={https://support.zoom.com/hc/en/article?id=zm_kb&sysparm_article=KB0059762},
  note={Accessed: 2025-11-21}
}

@misc{otter_ai,
  author={Otter.ai Inc.},
  title={Otter Meeting Agent - AI Notetaker, Transcription, Insights},
  year={2025},
  url={https://otter.ai},
  note={Accessed: 2025-11-21}
}

@misc{liu_listening_2025,
    title = {Listening, {Imagining} \& {Refining}: {A} {Heuristic} {Optimized} {ASR} {Correction} {Framework} with {LLMs}},
    shorttitle = {Listening, {Imagining} \& {Refining}},
    url = {http://arxiv.org/abs/2509.15095},
    doi = {10.48550/arXiv.2509.15095},
    language = {en},
    urldate = {2026-04-17},
    publisher = {arXiv},
    author = {Liu, Yutong and Zhang, Ziyue and Huang, Cheng and Yu, Yongbin and Wang, Xiangxiang and Cai, Yuqing and Tashi, Nyima},
    month = sep,
    year = {2025},
    note = {arXiv:2509.15095 [eess]},
}

\end{document}